\documentstyle[12pt,epsf,axodraw]{article}
\begin{document}
\begin{flushright}
{\bf LAPTH-936/02}
\end{flushright}
\vspace{1.0cm}

\begin{center}
{\Large \bf Triple and Quartic Interactions of Higgs Bosons in the
                 Two-Higgs-Doublet Model with 
                     CP violation \hfill\\}
\end{center}
\vspace{0.5cm}

\begin{center}
{ M.N.Dubinin$^{1}$, A.V.Semenov$^2$ \\
\vspace{5mm}
       {\small \it $^1$Institute of Nuclear Physics, Moscow State
University}\\  
       {\small \it  119992 Moscow, Russia} \\ 
       {\small \it $^2$Laboratoire de Physique Theorique }
         {LAPTH}\\
       {\small \it Chemin de Bellevue, B.P. 110, F-74941,
Annecy-le-Vieux, Cedex, France} }
\end{center}

\vspace{1.0cm}
\begin{center}
{\bf Abstract}
\end{center}
\begin{quote}
We consider the two-Higgs-doublet model with explicit
CP vio\-lation, where the effective Higgs potential  
is not CP invariant at the tree-level.
Three neutral Higgs bosons of the model are the mixtures
of CP-even and CP-odd bosons which exist in the
CP conserving limit of the theory. The mass spectrum and  
tree-level couplings of the neutral Higgs bosons to 
gauge bosons and fermions are significantly 
dependent on the parameters of the Higgs boson mixing matrix.
We calculate the Higgs-gauge boson, Higgs-fermion,
triple and quartic Higgs self-interactions in the 
MSSM with explicit CP violation in the Higgs sector  
and CP violating Yukawa interactions of the third generation 
scalar quarks. In some regions of the MSSM parameter space
substantial changes of the self-interaction vertices
take place, leading to significant suppression or enhancement of
the multiple Higgs boson production cross sections.
\end{quote}

\newpage

\section{Introduction}

General interest to the models with two (and more) Higgs doublets is 
maintained by the absence of a convincing argument in favor of only one
generation of Higgs bosons when there are three known generations of
fundamental fermions. 
Models with extended Higgs sector provide richer physical possibilities
than the standard scheme with one doublet. 
One of them is the possibility to introduce CP violation beyond the 
Cabibbo-Kobayashi-Maskawa (CKM) mechanism, by means of the Higgs boson 
exchange amplitudes with complex Higgs boson-fermion vertices.
Complex couplings can be generated either spontaneously \cite{Lee}, when
the vacuum expectation values of the Higgs fields are complex and  
couplings of the CP invariant tree-level Higgs potential are real, or 
explicitly inserted \cite{Georgi} on the level of 
$SU(2)\times U(1)$-invariant potential
terms, when the complex vacuum expectation values of scalar fields 
correspond to the minimum of hermitian potential with complex 
couplings, which is not CP-invariant
(CP-invariance softly broken by the mass terms). 

Various representations of the $SU(2)\times U(1)$-invariant 
two-doublet Higgs potentials
have been considered in the literature. The two-doublet models with 
spontaneous CP violation \cite{Lee,spontaneous} make use of the 
potential of general structure $-\mu^2 \varphi^2+\lambda \varphi^4$  
without the dimension two $\mu^2_{12}$-terms. 
Models with explicit CP violation use either the potential 
with trivial minimization (\cite{Georgi}, see also \cite{HHG}) or   
the potential with complex coupling $\mu^2_{12}$ of
the dimension two terms and complex couplings
$\bar \lambda_5$, $\bar \lambda_6$ and $\bar \lambda_7$ in front of the 
dimension
four potential terms \cite{PilaftsisWagner, Ellis}, similar to 
effective potential of the minimal supesymmetry (MSSM).
Standard transformation (diagonalization) procedure from the
level of primary fields which are the components of scalar 
doublets in $SU(2)\times U(1)$-invariant potential terms 
($SU(2)\times U(1)$ eigenstates), to the 
physical fields (mass eigenstates) of Higgs bosons should be 
consistently 
performed to respect the $SU(2)\times U(1)$ invariance and the 
minimization of the potential. We consider the diagonalization
for the two different two-Higgs-doublet potential forms in much 
details. 
A special case of the general two-Higgs-doublet model is represented by 
the Higgs sector of MSSM. Substantial radiative corrections to the Higgs 
boson masses and couplings are induced at the $m_Z$ scale mainly by the 
third generation quarks $t$, $b$ and the third generation scalar 
quarks \cite{RC}. In the special case of MSSM the multiparameter space 
of general two-Higgs-doublet model is significantly reduced, providing
possibilities of much less ambiguous phenomenological predictions.

Phenomenological consequences of the CP violating Higgs-third 
generation squark Yukawa interactions in the Higgs-fermion
and the Higgs-gauge boson sectors have been considered in 
\cite{PilaftsisWagner}. We focus mainly on the self-interactions of 
Higgs bosons. Experimental observation of the scalar boson signals
should be followed by the verification of Higgs mechanism
as the essence of the gauge boson and fermion mass generation.
Self-interactions of the Higgs fields lead to untrivial structure
of the vacuum state with nonzero (and possibly complex) field tensions,
initializing the spontaneous breakdown of $SU(2)\times U(1)$
symmetry.
Reconstruction of the Higgs self-interaction potential from the
data on multiple (mainly double and triple) Higgs boson production cross 
sections \cite{reconstruction}
requires the experimental
measurements of triple and quartic Higgs boson self-interaction
vertices, which is nontrivial but valuable task for a future high 
luminosity colliders, such as LHC and TESLA.

In section 2 we discuss the diagonalization of CP invariant Higgs 
potential, represented in two different forms, in the general 
two-Higgs-doublet model (THDM) and consider the special case
of the Higgs sector in the minimal supersymmetry (MSSM).
In section 3 we introduce complex parameters 
of the $SU(2)\times U(1)$ invariant potential terms and discuss
the diagonalization of the THDM potential which acquires explicit CP 
violation. In sections 4 and 5 we calculate 
the Higgs-gauge boson, Higgs-fermion and Higgs self-couplings in the 
MSSM with CP violation.

\section{Diagonalization of the mass matrix in the general
         two-Higgs-doublet model}

Two representations have been used for the two-doublet Higgs potential. 
The first representation \cite{Georgi,HHG}
\begin{eqnarray}
V(\varphi_1,\varphi_2)=&
         \lambda_1 (\varphi_1^+ \varphi_1 -\frac{v_1^2}{2})^2
        +\lambda_2 (\varphi_2^+ \varphi_2 -\frac{v_2^2}{2})^2 \\ \nonumber
       & +\lambda_3 [(\varphi_1^+ \varphi_1 -\frac{v_1^2}{2})
                  +(\varphi_2^+ \varphi_2 -\frac{v_2^2}{2})]^2 \\
\nonumber
       & +\lambda_4 [(\varphi_1^+ \varphi_1)(\varphi_2^+ \varphi_2)
                  -(\varphi_1^+ \varphi_2)(\varphi_2^+ \varphi_1)] \\ \nonumber
 &  +\lambda_5[{\tt Re}(\varphi_1^+ \varphi_2)-\frac{v_1 v_2}{2} 
               {\tt Re}(e^{i\xi})]^2
   +\lambda_6[{\tt Im}(\varphi_1^+ \varphi_2)-\frac{v_1 v_2}{2} 
              {\tt Im}(e^{i\xi})]^2
\end{eqnarray}
where $\lambda_i$ are real constants and the $SU(2)$ doublets 
$\varphi_{1,2}$ have the components
\begin{equation}
\varphi_1 = \{ -iw^+_1, \frac{1}{\sqrt{2}}(v_1+h_1+iz_1)\}, \quad
\varphi_2 = \{-iw^+_2, \frac{1}{\sqrt{2}}(v_2+h_2+iz_2)\}.
\end{equation}
$w_{1,2}$ are complex fields and $z_{1,2}$, $h_{1,2}$ are real scalar 
fields.
At positive $\lambda_1,...\lambda_6$ each term of the potential
$V(\varphi_1,\varphi_2)$ is obviously positive and its zero minimum 
is achieved if the vacuum expectation values of $<\varphi_1>$,
$<\varphi_2>$ are taken in the form 
\begin{equation}
<\varphi_1>= \frac{1}{\sqrt{2}} \{ 0, v_1 \}, \quad
<\varphi_2>= \frac{1}{\sqrt{2}} \{ 0, v_2 e^{i \xi} \}
\end{equation}
In the case of $\lambda_5=\lambda_6$ (corresponding to the
CP conserving MSSM-like potential, see below) the last two terms in (1)
form the modulo squared, and the phase $\xi$ can be removed from the
potential by the $U(1)$ rotation of $\varphi_2$, that does not affect
minimization. 
In this section we will consider the case $\lambda_5 
\neq \lambda_6$, $\xi=$0.  
Substitution of (2) to (1) gives a bilinear form of the mass term with
mixed components $w_i,h_i,z_i$, which can be diagonalized by an orthogonal
transformation of the fields in order to define the tree level masses of
Higgs bosons. 
In the CP conserving case the potential terms involving $z_1, z_2$
fields from the imaginary parts of $\varphi_1$, $\varphi_2$ doublets and 
$h_1, 
h_2$ fields from their real parts 
do not mix, so the mass terms are diagonalized by a separate 
two-dimensional rotations of the $z_1, z_2$ and the $h_1, h_2$ fields.
The resulting spectrum of scalars consists of two charged
$H^\pm$, three neutral $h$, $H$, $A^0$ scalar fields, and three Goldstone
bosons $G$. This procedure is described in many papers (for instance,
\cite{HHG, GHI}).
The $w_{1,2}$ sector is diagonalized by the rotation of
$w_1, w_2 \rightarrow H,G$
\begin{equation}
w_1^\pm =-H^\pm s_\beta +G^\pm c_\beta, \quad
w_2^\pm = H^\pm c_\beta + G^\pm s_\beta
\end{equation} 
defined by the angle
\begin{equation}
{\tt tg} \beta = \frac{v_2}{v_1}
\end{equation}
and leading to the massless $G$ field and the field of massive charged Higgs
boson $H^\pm$, $m_{H\pm}^2=\lambda_ 4(v_1^2+v_2^2)/2$. 
The $z_{1,2}$ sector is diagonalized by the rotation 
$z_1,z_2 \rightarrow A^0, G^{'}$ defined by
the angle $\beta$ and giving again one massless field $G^{'}$ and the 
field of CP-odd Higgs boson $A^0$ with the mass 
$m_{A}^2=\lambda_6(v_1^2+v_2^2)/2$.
Finally, the $h_1,h_2$ sector
is diagonalized by the rotation $h_1, h_2 \rightarrow h, H$ defined by the
angle $\alpha$ 
\begin{equation}
{\tt sin} 2\alpha =\frac{2m_{12}}{\sqrt{(m_{11}-m_{22})^2+4m_{12}^2}},
\quad
{\tt cos} 2\alpha
=\frac{m_{11}-m_{22}}{\sqrt{(m_{11}-m_{22})^2+4m_{12}^2}}
\end{equation}
where
\begin{eqnarray*}
m_{11}=&\frac{1}{4}[4v_1^2(\lambda_1+\lambda_3)+v_2^2\lambda_5] \\ \nonumber
m_{22}=&\frac{1}{4}[4v_2^2(\lambda_2+\lambda_3)+v_1^2\lambda_5] \\
\nonumber
m_{12}=&\frac{1}{4}(4\lambda_3+\lambda_5) v_1 v_2
\end{eqnarray*}    
giving two massive fields of CP-even Higgs bosons $H,h$ with masses
\begin{equation}
m^2_{H,h}=m_{11}+m_{22} \pm \sqrt{(m_{11}-m_{22})^2+4m_{12}^2}
\end{equation}
The diagonal mass term of scalar fields and their
triple and quartic self-interaction vertices can be 
explicitly obtained
by the substitution of the following expressions for $\lambda_{i}$ to the
potential $V(\varphi_1, \varphi_2)$ (1):
\begin{eqnarray}
\lambda_1=&\frac{1}{2v^2} \frac{1}{c_{\beta}^2}
         [\frac{s_{\alpha}}{s_{\beta}} c_{\alpha-\beta} m_h^2
        - \frac{c_{\alpha}}{s_{\beta}} s_{\alpha-\beta} m_H^2]
        + \frac{c_{2\beta}}{4c^2_{\beta}} \lambda_5  \\ \nonumber
\lambda_2=&\frac{1}{2v^2} \frac{1}{s_{\beta}^2}
         [\frac{c_{\alpha}}{c_{\beta}} c_{\alpha-\beta} m_h^2
        + \frac{s_{\alpha}}{c_{\beta}} s_{\alpha-\beta} m_H^2]
        - \frac{c_{2\beta}}{4s^2_{\beta}} \lambda_5  \\ \nonumber
\lambda_3=&\frac{1}{2v^2}[-\frac{s_{2\alpha}}{s_{2\beta}} m_h^2
                  + \frac{s_{2\alpha}}{s_{2\beta}} m_H^2]  
                  - \frac{1}{4}\lambda_5  \\ \nonumber
\lambda_4=&\frac{2}{v^2}m_{H^\pm}^2 \\        \nonumber
\lambda_6=&\frac{2}{v^2}m_{A^0}^2                        
\end{eqnarray} 
where we used the notation $v^2=v_1^2+v_2^2$, $s_{\alpha}={\tt sin}
\alpha$,
$c_\alpha= {\tt cos} \alpha$. Diagonalization of the mass term
takes place for arbitrary $\lambda_5$, which is a free parameter 
of the model. 

The second representation of the Higgs potential   
\begin{eqnarray}
U(\varphi_1, \varphi_2)=& -\mu_1^2 (\varphi_1^+ \varphi_1)
                         -\mu_2^2 (\varphi_2^+ \varphi_2)
       -\mu_{12}^2 (\varphi_1^+ \varphi_2+ \varphi_2^+ \varphi_1) \\  \nonumber     
     &+\bar \lambda_1 (\varphi_1^+ \varphi_1)^2
     +\bar \lambda_2 (\varphi_2^+ \varphi_2)^2
   +\bar \lambda_3 (\varphi_1^+ \varphi_1)(\varphi_2^+ \varphi_2) \\  \nonumber
   &+\bar \lambda_4 (\varphi_1^+ \varphi_2)(\varphi_2^+ \varphi_1)
    +\frac{\bar \lambda_5}{2} 
       [(\varphi_1^+ \varphi_2)(\varphi_1^+\varphi_2)
          + (\varphi_2^+ \varphi_1)(\varphi_2^+ \varphi_1)]
\end{eqnarray}
originates from the general SUSY action after the 
integration over Grassman variables and introduction of the soft 
SUSY-breaking terms (see \cite{HHG}).  
It is easy to check that 
the potentials (1) and (9) are equivalent if parameters 
$\bar \lambda_i$, $\mu^2_1,\, \mu^2_2,\, \mu^2_{12}$ and $\lambda_i$ are 
related by 
the formulae
\begin{eqnarray}
\bar \lambda_1= \lambda_1+\lambda_3, \quad
\bar \lambda_2= \lambda_2+\lambda_3, \quad
\bar \lambda_3= 2\lambda_3+\lambda_4, \quad    \\ \nonumber
\bar \lambda_4= -\lambda_4+ \frac{\lambda_5}{2}+\frac{\lambda_6}{2}, \quad
\bar \lambda_5= \frac{\lambda_5}{2}-\frac{\lambda_6}{2}, \quad 
\mu_{12}^2=\lambda_5 \frac{v_1 v_2}{2}
\end{eqnarray}
and
\begin{equation}
\mu_1^2=\lambda_1 v_1^2+\lambda_3 v_1^2 +\lambda_3 v_2^2, \quad
\mu_2^2=\lambda_2 v_2^2+\lambda_3 v_1^2 +\lambda_3 v_2^2
\end{equation}       
Unlike the potential (1) where the minimization is obvious, the 
symbolic structure of (9) does not demonstrate evidently its minimum.
The substitution of (2) to (9) gives linear terms in the component
fields $z_{1,2}$, $h_{1,2}$ (or physical fields $h,H,A$) and 
unless some additional conditions to
remove the linear terms are imposed, we are not in the minimum of
the potential. So the equations (11) which set to zero
the terms which are linear in component fields are  
the minimization conditions. The diagonalization of
$U(\varphi_1,\varphi_2)$ takes place for arbitrary $\mu^2_{12}$ 
parameter.  

Inverse transformation (10) has
the form
\begin{eqnarray}
\lambda_1=\bar \lambda_1 - \frac{\bar \lambda_3}{2}
                         - \frac{\bar \lambda_4}{2}
                         - \frac{\bar \lambda_5}{2} + \frac{\lambda_5}{2},
\hspace{4mm}
\lambda_2=\bar \lambda_2 - \frac{\bar \lambda_3}{2}
                         - \frac{\bar \lambda_4}{2}
                         - \frac{\bar \lambda_5}{2} + \frac{\lambda_5}{2}
\\ \nonumber
\lambda_3=                 \frac{\bar \lambda_3}{2}
                         + \frac{\bar \lambda_4}{2}
                         + \frac{\bar \lambda_5}{2} - \frac{\lambda_5}{2},
\hspace{4mm}
\lambda_4= -{\bar \lambda_4}-{\bar \lambda_5}+\lambda_5, \hspace{4mm}
\lambda_6= -2 \, {\bar \lambda_5} + \lambda_5 \nonumber
\end{eqnarray}
so masses of the CP-even scalars and their mixing
angle $\alpha$ (6),(7) in the case of potential $U(\varphi_1,
\varphi_2)$ can be easily obtained using
\begin{eqnarray}
\hspace{-10mm}
m_{11} +m_{22}=
v^2_1 \bar \lambda_1 +v^2_2 \bar \lambda_2 +\frac{\mu^2_{12}}{s_{2\beta}},
&& \hspace{-8mm}
m_{11}-m_{22}= v^2_1\bar \lambda_1 -v^2_2\bar \lambda_2 -{\tt
ctg}\, 2\beta \mu^2_{12},\\ \nonumber
2 m_{12} &=& v_1 v_2 (\bar \lambda_3+\bar \lambda_4+\bar \lambda_5)
- \mu^2_{12}. \nonumber
\end{eqnarray}
 
The diagonal form of $U(\varphi_1,\varphi_2)$ and the physical scalar
boson self-interaction vertices are obtained by substitution of the
following expressions for $\bar \lambda_{i}$ and $\mu_i$ to (9):
\begin{eqnarray}
\bar \lambda_1=&\frac{1}{2v^2} 
         [(\frac{s_{\alpha}}{c_{\beta}})^2 m_h^2
        + (\frac{c_{\alpha}}{c_{\beta}})^2 m_H^2  
     -  \frac{s_{\beta}}{c^3_{\beta}}\mu_{12}^2  ] \\ 
\bar \lambda_2=&\frac{1}{2v^2} 
         [(\frac{c_{\alpha}}{s_{\beta}})^2 m_h^2
        + (\frac{s_{\alpha}}{s_{\beta}})^2 m_H^2
     -  \frac{c_{\beta}}{s^3_{\beta}}\mu_{12}^2  ] \\ 
\bar \lambda_3=&\frac{1}{v^2}[2m^2_{H^\pm}   
                - \frac{\mu_{12}^2}{s_{\beta} c_{\beta}}
              +\frac{s_{2\alpha}}{s_{2\beta}} (m_H^2-m_h^2)]\\ 
\bar \lambda_4=&\frac{1}{v^2}(\frac{\mu_{12}^2}{s_{\beta} c_{\beta}} 
                   +m^2_A- 2 m^2_{H^\pm} ) \\ 
\bar \lambda_5=&\frac{1}{v^2} (\frac{\mu_{12}^2}{s_{\beta} c_{\beta}}
                    -m^2_{A} )               \\ 
\mu^2_1=&\bar \lambda_1 
v^2_1+(\bar \lambda_3+\bar \lambda_4+\bar \lambda_5)\frac{v^2_2}{2}
- \mu_{12}^2 {\tt tg}\beta \\ 
\mu^2_2=&\bar \lambda_2 
v^2_2+(\bar \lambda_3+\bar \lambda_4+\bar \lambda_5)\frac{v^2_1}{2}
- \mu_{12}^2 {\tt ctg}\beta 
%
%
\end{eqnarray}
Conditions (14)-(18) ensure the diagonal form of the mass term
expressed in physical fields $h, \, H, \, A, \, H^{\pm}$ 
and (19)-(20) are the minimization conditions.
Two parametrizations for the Higgs boson self-interaction vertices can be
used in THDM. In the first parametrisation \cite{DS} $\,\,$
$\mu^2_{12}$ is a free parameter and $\bar \lambda_5$ is defined by
(18). In the second one $\bar \lambda_5$ is a free
parameter and $\mu^2_{12}$ is equal to $s_{\beta} c_{\beta}(v^2 \bar
\lambda_5+m^2_A)$.
Complete sets of Feynman rules (unitary gauge) for the triple
($\mu^2_{12}$ and $\lambda_5$ parametrisations) and quartic ($\mu^2_{12}$
parametrisation) Higgs boson interactions in the general two-Higgs-doublet
model with CP-conservation are shown in Tables 1-2 
\footnote{These sets were obtained
by means of the LanHEP package \cite{LanHEP}, see
{\tt http://theory.sinp.msu.ru/\~{}semenov/lanhep.html}
Six misprints in sign of the second term in 
the factor $(c^3_\alpha c_\beta -s^3_\alpha s_\beta)$ that have taken 
place in
\cite{DS}, are corrected in Table 3 in the expressions for 
vertices $hhhh$, $AAAA$, $hhAA$, $H^+ H^- H^+ H^-$, $H^+ H^- AA$, $H^+ H^- 
h h$.}.
In the case of MSSM potential at the scale $M_{SUSY}$ (see (29)) $\bar 
\lambda_5=$0 and it follows from 
(8),(10),(11) that $\mu^2_{12}$ is fixed and equal to $m^2_A s_{\beta} 
c_{\beta}$.

Two additional terms of the dimension 4 can be constructed 
using the complete set of $SU(2)\times U(1)$ invariants 
$\varphi^+_1 \varphi_1$, $\varphi^+_2 \varphi_2$, 
{\tt Re}$\varphi^+_1 \varphi_2$ and {\tt Im}$\varphi^+_1 \varphi_2$
(a detailed discussion of all possible potental forms can 
be found in \cite{SantosBarroso}). These
terms are usually added to the $U(\varphi_1, \varphi_2)$ with the 
couplings $\bar \lambda_6$ and $\bar \lambda_7$ 
\begin{eqnarray}
{\bar U}(\varphi_1, \varphi_2)&=&U(\varphi_1, \varphi_2)\\ \nonumber
&&+\bar \lambda_6 (\varphi^+_1 \varphi_1) 
[(\varphi^+_1 \varphi_2)+(\varphi^+_2 \varphi_1)] 
+\bar \lambda_7 (\varphi^+_2 \varphi_2)
[(\varphi^+_1 \varphi_2)+(\varphi^+_2 \varphi_1)]
\end{eqnarray}
The diagonal form of ${\bar U}(\varphi_1, \varphi_2)$ at the 
local minimum takes place at arbitrary $\mu^2_{12}$, $\bar \lambda_6$, 
$\bar \lambda_7$ and can be
achieved by means of the substitution (14)-(20) with additional $\bar 
\lambda_6$, 
$\bar 
\lambda_7$ terms in the right-hand side:
\begin{eqnarray}
\bar \lambda_1& =& 
\frac{1}{2v^2}
         [(\frac{s_{\alpha}}{c_{\beta}})^2 m_h^2
        + (\frac{c_{\alpha}}{c_{\beta}})^2 m_H^2
     -  \frac{s_{\beta}}{c^3_{\beta}}\mu_{12}^2  ]
+\frac{1}{4}( \bar \lambda_7 {\tt tg}^3 {\beta} 
-3 \bar \lambda_6{\tt tg} {\beta}) \\ 
\bar \lambda_2& =&
          \frac{1}{2v^2}
         [(\frac{c_{\alpha}}{s_{\beta}})^2 m_h^2
        + (\frac{s_{\alpha}}{s_{\beta}})^2 m_H^2
     -  \frac{c_{\beta}}{s^3_{\beta}}\mu_{12}^2  ]
+\frac{1}{4}(\bar \lambda_6 {\tt ctg}^3 {\beta}  
-3 \bar \lambda_7 {\tt ctg} {\beta}) \\ 
\bar \lambda_3& =& 
\frac{1}{v^2}[2m^2_{H^\pm}
                - \frac{\mu_{12}^2}{s_{\beta} c_{\beta}}  
              +\frac{s_{2\alpha}}{s_{2\beta}} (m_H^2-m_h^2)]
-\frac{\bar \lambda_6}{2} {\tt ctg}\beta 
- \frac{\bar \lambda_7}{2} {\tt tg}\beta  \\ 
\bar \lambda_4& =& 
\frac{1}{v^2}(\frac{\mu_{12}^2}{s_{\beta} c_{\beta}}
                   +m^2_A- 2 m^2_{H^\pm} )
-\frac{\bar \lambda_6}{2} {\tt ctg}\beta  - 
\frac{\bar \lambda_7}{2} {\tt tg}\beta  \\
\bar \lambda_5& = &
\frac{1}{v^2} (\frac{\mu_{12}^2}{s_{\beta} c_{\beta}}
                    -m^2_{A} )
-\frac{\bar \lambda_6}{2} {\tt ctg}\beta - \frac{\bar \lambda_7}{2} {\tt 
tg}\beta \\ 
\mu^2_{1}& = &
\bar \lambda_1 v^2_1+(\bar \lambda_3+\bar 
\lambda_4+\bar \lambda_5)\frac{v^2_2}{2}
- \mu_{12}^2 {\tt tg}\beta
+\frac{v^2 s^2_{\beta}}{2}
(3 \bar \lambda_6 {\tt ctg} \beta + \bar \lambda_7 {\tt tg} \beta)\\ 
\mu^2_{2}& = &  {\hskip -2mm}
\bar \lambda_2 v^2_2+(\bar \lambda_3+\bar 
\lambda_4+\bar \lambda_5)\frac{v^2_1}{2}
- \mu_{12}^2 {\tt ctg}\beta 
+\frac{v^2 c^2_{\beta}}{2}              
( \bar \lambda_6 {\tt ctg}\beta + 3  \bar \lambda_7 {\tt tg} \beta )
\end{eqnarray}
Our expressions for the redefined $\bar \lambda_4$ and $\bar \lambda_5$ 
are the same as given in \cite{HaberHempfling}.

The potentials (1) and (9) can be reduced to the MSSM potential
in some regions of the parameter space which we are going to discuss.
The potential $V(\varphi_1, \varphi_2)$ (1) has eight
parameters: two vev's  $v_1$, $v_2$ and six couplings $\lambda_i$ 
($i$=1,...6).
Eight parameters of the potential $U(\varphi_1, \varphi_2)$ (9)
$\mu_1$, $\mu_2$, $\mu_{12}$ and $\bar \lambda_{i}$ ($i$=1,...5) can be
found using (10),(11). From the other side, in the
Higgs sector we have eight physical parameters:
the mixing angle $\beta$ and $W$-boson mass $m_W$, mixing angle
$\alpha$, the parameter $\mu^2_{12}$ and four masses of
scalars $m_h$, $m_H$, $m_A$, $m_{H^\pm}$. The $m_W$ is fixed 
experimentally
maintaining the constraint on the $v_1, \, v_2$, $\;$ 
$v^2=v^2_1+v^2_2={4 m^2_W}/{e^2}\cdot{\tt sin}^2 \theta_W$
which follows from the Higgs kinetic term
$D_{\mu}\varphi D^{\mu}\varphi$ ($g=e/{\tt sin} \theta_W$, $\theta_W$
is the Weinberg angle). So the Higgs sector of THDM with the potentials 
(1) or (9) 
is described by a seven-dimensional parameter space. In 
the case of superpotential five additional constraints are imposed,
relating all Higgs boson self-couplings $\bar \lambda_{i}$, ($i$=1,...5)
to the gauge coupling constants at the energy scale $M_{SUSY}$
\cite{Inoue}:
\begin{equation}
\bar \lambda^{SUSY}_1= \bar \lambda^{SUSY}_2=\frac{g^2+g^2_1}{8}, \;
\bar \lambda^{SUSY}_3=\frac{g^2-g^2_1}{4},           \;
\bar \lambda^{SUSY}_4=-\frac{g^2}{2},                \;
\bar \lambda^{SUSY}_5=0.
\end{equation}
It follows that the four Higgs boson masses and the two mixing angles are 
defined by two independent parameters. One can
choose, for example, the $r_1, r_2$ parametrization \cite{GHII}
($r_{1,2}=m^2_{h,H}/m^2_Z$) or the well-known $m_A$, ${\tt tg} \beta$
parametrization. In order to reduce the general two-Higgs-doublet model
vertices to MSSM at the scale $M_{SUSY}$ it is convenient to use the
$\alpha$, $\beta$ parametrization:
\begin{equation}
m^2_h=m^2_Z 
c_{2\beta} \, \frac{s_{\alpha+\beta}}{
s_{\alpha-\beta}},
\quad
m^2_H=m^2_Z 
c_{2\beta} \, \frac{c_{\alpha+\beta}}{
c_{\alpha-\beta}},
\quad
m^2_A=m^2_Z \frac{s_{2(\alpha+\beta)}}{s_{2(\alpha-\beta)}},
\quad
\mu^2_{12}= m^2_A s_{\beta} c_{\beta}.
\end{equation}
Substitution of these expressions to the vertex factors in Tables 1,2
after trivial trigonometric transformations reduces
them \cite{DS} to a simpler MSSM factors (see \cite{HHG}). However, (30) 
are no longer valid at the energy scale $m_W$ where $\bar 
\lambda^{SUSY}_i$ 
couplings and masses of Higgs bosons are 
significantly changed by radiative corrections and the effective 
two-Higgs-doublet potential should be described in the  
complete seven dimensional parameter space. 
Practical calculations of the radiatively corrected masses and/or 
couplings
can be conveniently carried out using
results of the two approaches, renormalization group (the HMSUSY 
package \cite{HHH} or the analytical representation \cite{twoloop}) and 
diagrammatic (the FeynHiggsFast package, see \cite{diag}). 
Two different parametrizations can be used for these approaches.

In the RG approach it seems convenient to use the two-Higgs-doublet model 
parameter space 
$m_A$, ${\tt tg}\beta$, $\bar \lambda_1,...\bar \lambda_5$.
In the following we shall take into account the $\bar \lambda_6$ and
$\bar \lambda_7$ terms defined by (21), so the parameter space
will be nine-dimensional.    
RG evolution of the coupling constants $\bar \lambda_i$
from the energy scale $M_{SUSY}$ to the electroweak scale $m_W$ 
defines the $\bar \lambda_1$,...$\bar \lambda_5$ in (22)-(26) and the 
parameters $\bar \lambda_6$, $\bar \lambda_7$. At a given $m_A$, ${\tt 
tg}\beta$,
$\bar \lambda_6$, $\bar \lambda_7$ the parameter $\mu^2_{12}$ and 
$m_{H^\pm}$ are 
fixed by the conditions (25) and (26),
the parameters $\mu^2_1$ and $\mu^2_2$ are fixed by (27) and (28), 
$\alpha$, 
$m_h$ and $m_H$ can be found using the equations (22)-(24).
If we denote the deviation from the coupling $\bar \lambda^{SUSY}_i$ at 
the MSSM scale by $\Delta \bar \lambda_i$
\begin{eqnarray*}
2(\bar \lambda^{SUSY}_{1,2} - \bar \lambda_{1,2}) = \Delta \bar 
\lambda_{1,2}, \quad
\bar \lambda^{SUSY}_{3,4} - \bar \lambda_{3,4} = \Delta \bar 
\lambda_{3,4}, \quad
-\bar \lambda_{5,6,7}= \Delta \bar \lambda_{5,6,7}
\end{eqnarray*}
we find the mixing angle (introducing the notation 
$g^2_1+g^2=g^2\, 
m^2_Z/m^2_W$, $g^2-g^2_1=g^2(2-\, m^2_Z/m^2_W)$ while using (22)-(26)) 
\begin{eqnarray}
{\tt tg} 2\alpha &\hspace{-3mm} =& \hspace{-5mm}
\frac { s_{2\beta} (m^2_A + m^2_Z)  +v^2 ((\Delta
\bar \lambda_3 + \Delta \bar \lambda_4) s_{2\beta}+2c^2_{\beta} 
\Delta \bar \lambda_6
+ 2s^2_{\beta} \Delta \bar \lambda_7) }
                        {c_{2\beta}(m^2_A - m^2_Z) +v^2 (\Delta \bar
\lambda_1 c^2_{\beta} - \Delta \bar \lambda_2 s^2_{\beta} - \Delta \bar 
\lambda_5
c_{2\beta}+(\Delta \bar \lambda_6- \Delta \bar \lambda_7)s_{2\beta} )}  
\end{eqnarray}
CP-even Higgs boson masses and the $\mu^2_{12}$ parameter
\begin{eqnarray}
m^2_H&=&c^2_{\alpha+\beta} m^2_Z + s^2_{\alpha-\beta} m^2_A \\ \nonumber
&&\hspace{-16mm}
 -v^2 (\Delta \bar \lambda_1 c^2_{\alpha} c^2_{\beta} +\Delta 
\bar \lambda_2
s^2_{\alpha} s^2_{\beta}+ 2(\Delta \bar \lambda_3+\Delta 
\bar \lambda_4)c_{\alpha}
c_{\beta} s_{\alpha} s_{\beta}+ \Delta \bar \lambda_5 (c^2_{\alpha} 
s^2_{\beta}
+s^2_{\alpha} c^2_{\beta})) \\ \nonumber
&&
+2 s_{\alpha+\beta} (\Delta \bar \lambda_6 c_{\alpha} c_{\beta}
                    +\Delta \bar \lambda_7 s_{\alpha} s_{\beta}) \\ 
\nonumber
m^2_h&=&s^2_{\alpha+\beta} m^2_Z + c^2_{\alpha-\beta} m^2_A \\ \nonumber
&&\hspace{-16mm}
- v^2 (\Delta \bar \lambda_1 s^2_{\alpha} c^2_{\beta} +\Delta 
\bar \lambda_2
c^2_{\alpha} s^2_{\beta}- 2(\Delta \bar \lambda_3+\Delta 
\bar \lambda_4)c_{\alpha}
c_{\beta} s_{\alpha} s_{\beta}+ \Delta \bar \lambda_5 (s^2_{\alpha} 
s^2_{\beta}
+c^2_{\alpha} c^2_{\beta})) \\ \nonumber
&&
-2 c_{\alpha+\beta} (\Delta \bar \lambda_6 s_{\alpha} c_{\beta}
                    -\Delta \bar \lambda_7 c_{\alpha} s_{\beta}) \\
m^2_{H^\pm}&=&m^2_W+m^2_A-\frac{v^2}{2}(\Delta \bar \lambda_5-\Delta \bar 
\lambda_4) \\
\nonumber
\mu^2_{12}&=&s_{\beta} c_{\beta}[m^2_A-\frac{v^2}{2}(2\Delta \bar
\lambda_5+\Delta \bar
\lambda_6
{\tt ctg} \beta+\Delta \bar \lambda_7 {\tt tg} \beta)] 
\nonumber
\end{eqnarray}
with the minimization conditions
\begin{eqnarray}
\mu^2_1&=&\frac{1}{2} m^2_Z c_{2\beta}-\mu^2_{12} {\tt tg}\beta  \\ 
\nonumber
&&-\frac{v^2}{2}[\Delta \bar \lambda_1 c^2_{\beta}+(\Delta \bar 
\lambda_3+\Delta \bar \lambda_4
+\Delta \bar \lambda_5) s^2_{\beta}
+3\Delta \bar \lambda_6 s_{\beta} c_{\beta}+ \Delta \bar \lambda_7 
\frac{s^3_{\beta}}{ c_{\beta}}] \\ \nonumber
\mu^2_2&=&-\frac{1}{2} m^2_Z c_{2\beta}-\mu^2_{12} {\tt ctg}\beta  
\\ \nonumber
&&-\frac{v^2}{2}[\Delta \bar \lambda_2 s^2_{\beta}+ (\Delta 
\bar \lambda_3+\Delta \bar \lambda_4
+\Delta \bar \lambda_5) c^2_{\beta}
+\Delta \bar \lambda_6 \frac{c^3_{\beta}}{ s_{\beta}}+ 3\Delta \bar 
\lambda_7 
s_{\beta} 
c_{\beta}] 
\end{eqnarray}
These expressions can be straighforwardly used to calculate the radiatively 
corrected masses of Higgs bosons and the mixing angle $\alpha$ in the 
MSSM with the help of a solution of the RG equations for 
$\bar \lambda_1,...\bar \lambda_7$. 
Apparently, in the RG 
approach Feynman rules in terms of $\bar \lambda_i$ couplings are more 
convenient than rules in terms of Higgs particle masses. 

In the diagrammatic approaches to calculation of the radiatively corrected 
masses \cite{diag} the corrections to $m_h$, $m_H$, $m_A$ and $m_{H^\pm}$ 
are extracted from the renormalized Higgs boson self-energies
(usually radiative corrections to
only the CP-even Higgs boson masses are calculated). The set of 7+2
independent parameters inherent for
the diagrammatic approaches could be $m_A$, ${\tt tg}\beta$, $\alpha$, 
$\mu^2_{12}$, $m_h$, 
$m_H$, $m_{H^\pm}$, and $\bar \lambda_6, \, \bar \lambda_7$. At a given
$m_A$, ${\tt tg}\beta$, $\bar \lambda_6, \, \bar \lambda_7$ the 
$\mu^2_{12}$ parameter can be fixed at the value  $m^2_A s_{\beta} 
c_{\beta}$, and $\alpha$ can be calculated using the renormalized 
self-energies correction \cite{diag} to the relation valid at the 
$M_{SUSY}$ scale $m^2_A+m^2_Z= -s_{2\alpha}/s_{2\beta}(m^2_H-m^2_h)$. 
Then $\bar \lambda_4$ is defined by (25) and 
$\bar \lambda_1$,...$\bar \lambda_3$ can be found using (22)-(24).
In the diagrammatic calculations Feynman rules in terms of the 
radiatively corrected Higgs boson masses look
more natural. Substitution of the radiatively corrected Higgs 
masses to the tree-level Higgs vertex factors is expected to give results
very close to those obtained from the loop corrections to Higgs
vertex at the SUSY scale (see the discussion in the last
ref. \cite{reconstruction}). It has been shown in \cite{Hollik} 
on the example of $hhh$ and $hhhh$ 
vertices (and for the case of diagonal third generation squark mass 
matrix) that large radiative 
corrections to the vertex factors calculated diagrammatically
can be absorbed in the radiatively corrected Higgs boson masses. 
 
Other parametrizations in the two-Higgs-doublet model are of course 
possible, but they should be
carefully introduced to respect the minimization and 
diagonalization conditions (22)-(28). 
The introduction of scalar particle masses and mixing angles inconsistent
with them violates either diagonalizaton of the potential or its $SU(2)$ 
invariance, even if the minimization conditions remain valid.      

\section{CP violation in the two-Higgs-doublet model}

CP transformation of the scalar doublet $CP \varphi P^+ C^+ = 
\stackrel{*} \zeta_{CP} \varphi^+$ (the phase factor $|\zeta_{CP} |=$1)
changes the sign of
imaginary part ${\tt Im} (\varphi_1^+ \varphi_2)$ in the
$\lambda_6$ term of potential (1), so if $\lambda_5 \neq \lambda_6$,
$\lambda_6 \neq$0 and $\xi \neq$0, CP symmetry is 
broken there explicitly. In other words, the dimension
two terms of (1) appear with the complex parameter $\mu^2_{12}$
\begin{eqnarray}
\frac{\lambda_5}{4}[\varphi_1^+ \varphi_2 + \varphi_2^+ \varphi_1 - v_1
v_2 {\tt cos} \xi]^2 +
\frac{\lambda_6}{4}[-i(\varphi_1^+ \varphi_2 - \varphi_2^+ \varphi_1) -
v_1 v_2 {\tt sin} \xi]^2 \\ \nonumber
\Rightarrow (\frac{\lambda_5}{4}-\frac{\lambda_6}{4})[(\varphi_1^+
\varphi_2)^2
+(\varphi_2^+ \varphi_1)^2]
+(\frac{\lambda_5}{2}+\frac{\lambda_6}{2})
\, \varphi_1^+ \varphi_2 \, \varphi_2^+ \varphi_1 \\
\nonumber
-\frac{v_1 v_2}{2} (\lambda_5 {\tt cos} \xi- i \lambda_6 {\tt sin} \xi)
\, \varphi_1^+ \varphi_2
-\frac{v_1 v_2}{2} (\lambda_5 {\tt cos} \xi+ i \lambda_6 {\tt sin} \xi)
\, \varphi_2^+ \varphi_1
\end{eqnarray}  
so we find
\begin{eqnarray}
\mu^2_{12}=\frac{v_1 v_2}{2} (\lambda_5 {\tt cos} \xi- i \lambda_6 {\tt
sin} \xi).
\end{eqnarray}
CP is (softly) broken by the 
$\mu^2_{12} \varphi_1^+ \varphi_2+ 
\stackrel{*} \mu^2_{12} \varphi_2^+ \varphi_1$ terms.
In this special case when the same phase $\xi$ is involved
both in the potential and the vacuum expectation value of
$\varphi_2$, diagonalization and minimization of CP transformed
potential become less transparent.
It is more convenient to analyse
the potential form (21) for the general case of 
complex parameters with arbitrary phases.
In the following we shall consider the hermitian potential
which is the generalization of (21)
\begin{eqnarray}
\bar U(\varphi_1, \varphi_2)=&\frac{1}{2}[ -\mu_1^2 (\varphi_1^+ 
\varphi_1)
                -\stackrel{*}{\mu_1^2} (\varphi_1^+ \varphi_1)
                          -\mu_2^2 (\varphi_2^+ \varphi_2)
     -\stackrel{*}{\mu_2^2} (\varphi_2^+ \varphi_2)] \\ \nonumber
      & -\mu_{12}^2 (\varphi_1^+ \varphi_2)
       -\stackrel{*}{\mu_{12}^2} (\varphi_2^+ \varphi_1) \\ \nonumber
     &+\frac{1}{2}[\bar \lambda_1 (\varphi_1^+ \varphi_1)^2
      +\stackrel{*}{\bar \lambda_1} (\varphi_1^+ \varphi_1)^2
      +\bar \lambda_2 (\varphi_2^+ \varphi_2)^2
+\stackrel{*}{\bar \lambda_2} (\varphi_2^+ \varphi_2)^2 \\ \nonumber
   &+\bar \lambda_3 (\varphi_1^+ \varphi_1)(\varphi_2^+ \varphi_2)
+\stackrel{*}{\bar \lambda_3} (\varphi_1^+ \varphi_1)(\varphi_2^+
\varphi_2) \\ \nonumber
   &+\bar \lambda_4 (\varphi_1^+ \varphi_2)(\varphi_2^+ \varphi_1)
+\stackrel{*}{\bar \lambda_4} (\varphi_1^+ \varphi_2)(\varphi_2^+  
\varphi_1)] \\ \nonumber
   &+\frac{\bar \lambda_5}{2}
       (\varphi_1^+ \varphi_2)(\varphi_1^+\varphi_2)
+\frac{\stackrel{*}{\bar \lambda_5}}{2}
          (\varphi_2^+ \varphi_1)(\varphi_2^+ \varphi_1) \\ \nonumber
&+\bar \lambda_6 (\varphi^+_1 \varphi_1)(\varphi^+_1 \varphi_2)+
\stackrel{*}{\bar \lambda_6}(\varphi^+_1 \varphi_1)(\varphi^+_2 \varphi_1)
\\ \nonumber
&+\bar \lambda_7 (\varphi^+_2 \varphi_2)(\varphi^+_1 \varphi_2)
+\stackrel{*}{\bar \lambda_7}(\varphi^+_2 \varphi_2)(\varphi^+_2 
\varphi_1) \nonumber
\end{eqnarray}
$\stackrel{*}{a}$ denotes $a$ complex conjugated. 
The potential terms with complex parameters $\mu^2_{12}$ and 
$\bar \lambda_5, \, \bar \lambda_6, \, \bar \lambda_7$
explicitly violate CP invariance. 
With the help of $U(1)_Y$ hypercharge symmetry of the model
the phases $\theta_{\mu}$ and $\theta_{5,6,7}$ of complex
parameters $\mu^2_{12}$ and $\bar \lambda_{5,6,7}$
can be removed by the phase rotation of scalar 
doublets $\varphi_{1,2}$. In the case when $v_1$ in (3)
is taken real and positive, and $\eta$ is the overall phase 
of $\varphi_2$ doublet,
the conditions to remove explicit phases from the potential
are \cite{GunionHaber}
\begin{eqnarray}
&\theta_{\mu}-\eta=n_{\mu} \pi, \quad
 \theta_5-2 \eta= n_5 \pi, \quad
 \theta_{6,7} -\eta= n_{6,7} \pi&
\end{eqnarray}
where $n_{\mu}$, $n_i$ ($i=$5,6,7) are arbitrary integer numbers. 
Or equivalently, 
in terms of complex parameters the conditions for the absence of explicit
CP violation in the effective potential (37) are \cite{PilaftsisWagner}
\begin{eqnarray}
&{\tt Im}({\mu}^4_{12} \stackrel{*}{\bar \lambda_5})=0, \quad
{\tt Im}({\mu}^2_{12} \stackrel{*}{\bar \lambda_6})=0, \quad
{\tt Im}({\mu}^2_{12} \stackrel{*}{\bar \lambda_7})=0.&
\end{eqnarray} 
The phase of complex ${\mu}^2_{12}$ can be always
rotated away, so ${\mu}^2_{12}$ can be taken real. In the scenario
of fine-tuning for the phases, when the conditions (38) or (39) are 
satisfied,
$\bar \lambda_i$ are also real. Otherwise
$\bar \lambda_i$ will be redefined after a phase rotation, keeping
explicitly CP violating terms.
 
If the phases of ${\mu}^2_{12}$, $\bar \lambda_i$ are rotated away so 
there are no explicitly CP noninvariant potential terms, CP invariance
can be nevertheless broken
spontaneously. Using the convention when $v_1$ and $v_2$ in formula (3) 
are real and positive and selecting 
the $\xi$ phase-dependent terms of (37), which are of the form
$a {\tt cos}^2 \xi + b {\tt cos} \xi$, where
$a=\bar \lambda_5 v^2_1 v^2_2/2$ and 
$b=(\bar \lambda_6 v^2_1/2+\bar \lambda_7 v^2_2/2-\mu^2_{12})v_1 v_2$ 
($0 \leq \xi \leq \pi$)
one can find their minimum 
(see the appendix of \cite{GunionHaber}) at
${\bar \lambda_5} \geq 0$ and
\begin{eqnarray}
{\tt cos}\xi=\frac{\mu^2_{12} -\frac{v^2}{2} \bar \lambda_6 c^2_{\beta}
-\frac{v^2}{2} \bar \lambda_7 s^2_{\beta} }
{\bar \lambda_5 v^2 s_{\beta} c_{\beta}}
\end{eqnarray}
(the special case $\mu^2_{12}=$0 was found in 
\cite{Lee,spontaneous}). 
If $\mid \mu^2_{12} -\frac{v^2}{2} \bar \lambda_6 c^2_{\beta}
-\frac{v^2}{2} \bar \lambda_7 s^2_{\beta} \mid \geq \bar \lambda_5 v^2 
s_{\beta} 
c_{\beta}$ then the spontaneously CP violating extremum does not exist
and the minimum of the potential takes place at the endpoints
${\tt cos}\xi=\pm 1$. For $b>0$ the minimum is at $\xi= \pi$ 
and for $b<0$ the minimum is at $\xi= 0$. If $\bar \lambda_5 <$0,
the extremum inside the interval $|{\tt cos}\xi|< 1$ is the maximum,
so the local minima are at the endpoints $\xi= 0, \pi$.
The case $\xi= 0$ is reduced to the case $\xi= \pi$ by the change of
sign for $\bar \lambda_6, \bar \lambda_7, \mu^2_{12}$ (change of
sign for the $b$).
The special case of CP conservation takes place at
$\xi= \pi/2$ when $\mu^2_{12}=\frac{v^2}{2} \bar \lambda_6 c^2_{\beta}
+\frac{v^2}{2} \bar \lambda_7 s^2_{\beta}$. In 
this case of purely imaginary $<\varphi_2>$
\cite{Lin} our diagonalization procedure must be reconsidered. 
For instance, CP-even Higgs mass eigenstates are formed in this case by 
different orthogonal linear combinations
of real and imaginary parts of scalar doublets, ${\tt Re}\varphi_1$ and 
${\tt Im}\varphi_2$. 

The substitution of (26) to the no-extremum
condition $|\mu^2_{12} -\frac{v^2}{2} \bar \lambda_6 c^2_{\beta}
-\frac{v^2}{2} \bar \lambda_7 s^2_{\beta}| \geq \bar \lambda_5 v^2
s_{\beta} c_{\beta}$ gives $m^2_A \geq$0.
In the case $\bar \lambda_5 <$0 the absolute minimum of the potential
takes place at $\xi=$0 (but not $\xi=\pi$) if $\mu^2_{12} -\frac{v^2}{2} 
\bar \lambda_6 
c^2_{\beta}
-\frac{v^2}{2} \bar \lambda_7 s^2_{\beta} \geq 0$, which gives
in combination with (26) $m^2_A \geq \mid \bar \lambda_5 \mid v^2$.
Even if $\bar \lambda_5$ is of the order of 1, the latter condition is
valid when $m_A$ is in the mass range
of the order or greater than $v$. In the case of real parameters
${\mu}^2_{12}$, $\bar \lambda_i$ it is not straightforward to combine
spontaneous CP violation with our procedure of diagonalization.
In the case of complex parameters the situation may be changed.
The extremum conditions can be found from the study of the
fourth power equation with the coefficients depending on
real and imaginary parts of ${\mu}^2_{12}$, $\bar \lambda_i$.
Nevertheless we are not going to consider spontaneous CP violation 
further on. With real parameters ${\mu}^2_{12}$, $\bar 
\lambda_i$ and in the absence of spontaneous CP violation
the minimum of the potential (37) can be taken at $\xi=$0.
At the same time insofar as the physical motivation of the fine-tuning
conditions (38),(39) for CP conservation is not available,
in the following we consider the general case of
diagonalization and minimization of the two-Higgs-doublet
potential with complex parameters.
 
For the diagonalization of (37) in the ground state we used the 
ansatz (22)-(28) to be taken for real parts of parameters. 
The real part of
$\mu^2_{12}$ parameter is expressed through the real parts of $\bar 
\lambda_{5,6,7}$ using (26)  
\begin{eqnarray}
{\tt Re} \mu^2_{12}&=&
m^2_A s_{\beta} c_{\beta}+v^2 ( s_{\beta} c_{\beta}
{\tt Re} \bar \lambda_5
+\frac{1}{2} c^2_{\beta} \; {\tt Re} \bar \lambda_6
+\frac{1}{2} s^2_{\beta} \; {\tt Re} \bar \lambda_7) 
\end{eqnarray}
defining also the real parts of $\bar \lambda_{1,2,3,4}$ and
$\mu^2_1,\, \mu^2_2$ by means of (22)-(25) and (27),(28). 
The substitution
of complex $\mu_i$ and $\bar \lambda_i$ 
to the potential (37) 
leads to the linear term and the non-diagonal mass term which are
dependent on the imaginary parts of $\mu^2_{12}, \bar \lambda_i$
\begin{eqnarray}
\bar U(\varphi_1, \varphi_2)& =& c_0 A + c_1 hA + c_2 HA \\ \nonumber 
&&+ \frac{m^2_h}{2} H^2 + \frac{m^2_H}{2} H^2
+\frac{m^2_A}{2} A^2 + m^2_{H^\pm} {H^+ H^-} \\ \nonumber
&&+ {\tt \; third \; and \; fourth \; order \; terms \; in} \; h,H,A,H^\pm
\end{eqnarray}
where
\begin{eqnarray}
c_0 &=& -v \, {\tt Im} \mu^2_{12}+\frac{v^3}{2} s_{\beta} c_{\beta}
{\tt Im} \bar \lambda_5 + \frac{v^3}{2}(c^2_{\beta} {\tt Im} \bar \lambda_6 +
                s^2_{\beta} {\tt Im} \bar \lambda_7)
  \\
\nonumber
c_1&=&-s_{\alpha-\beta} \, {\tt Im} \mu^2_{12}
       +\frac{v^2}{4} (s_{2\beta} s_{\alpha-\beta}-2 c_{\alpha+\beta})
       {\tt Im} {\bar \lambda_5} \\ \nonumber
     && -\frac{v^2}{2}(s_{\beta} c_{\beta} c_{\alpha-\beta}-3 s_{\alpha} c_{\beta})
       {\tt Im} {\bar \lambda_6} 
      +\frac{v^2}{2}(s_{\beta} c_{\beta} c_{\alpha-\beta}-3 c_{\alpha} s_{\beta})
       {\tt Im} {\bar \lambda_7} \\ \nonumber
c_2&=&c_{\alpha-\beta} \, {\tt Im} \mu^2_{12}
       +\frac{v^2}{4} (c_{2\beta} s_{\alpha-\beta}-3 s_{\alpha+\beta})
       {\tt Im} {\bar \lambda_5} \\ \nonumber
     && -\frac{v^2}{2}(c^2_{\beta} c_{\alpha-\beta}+2 c_{\alpha} c_{\beta})
       {\tt Im} {\bar \lambda_6}     
      -\frac{v^2}{2}(s^2_{\beta} c_{\alpha-\beta}+2 s_{\alpha} s_{\beta})
       {\tt Im} {\bar \lambda_7}  \nonumber
\end{eqnarray}
In the case of CP conservation (38), (39) the linear and non-diagonal 
second order terms $hA$ and $HA$ do not appear because all imaginary 
parts of parameters can be removed. The linear term in $A$
demonstrates that after the introduction of complex couplings we can be
out of a local minimum of the potential $\bar U(\varphi_1, \varphi_2)$.
The minimization condition for the imaginary parts $c_0=$0 must be 
imposed.
If $\bar \lambda_i$, $i=$5,6,7 are zero 
(this occurs in THDM if we additionally introduce a global $U(1)_Q$ 
symmetry, softly broken by the dimension two $\mu^2_{12}$ term 
\cite{Pilaftsis0}),
the imaginary part of $\mu^2_{12}$ can be removed by
the phase rotation of $\varphi_2$, so the tree-level potential
is CP invariant.
In the general THDM with nonzero parameters
the phase rotation of $\varphi_2$ which removes ${\tt Im}\mu^2_{12}$ 
redefines $\bar \lambda_i$, $i=$5,6,7. These simple observations 
for the THDM potential isolated from any other physical
fields are no longer trivial if, keeping in mind the MSSM, we switch
on the interaction of $\varphi_1$ and $\varphi_2$ with scalar quarks.
CP invariance of the latter is (softly) broken by the dimension three
terms with the Higgs mixing parameter $\mu$ and the trilinear 
parameters $A_{t,b}$ of the form $f_{1,2} A_{b,t} \varphi^0_{1,2}
\tilde q^*_R \tilde q_L$, $\mu f_{2,1}  \varphi^0_{1,2} \tilde q^*_L 
\tilde 
q_R$ ($q= \tilde t,\tilde b$; $\, \varphi^0_{1,2}$ are the neutral 
components of the Higgs doublets, $f_{1,2}=\sqrt{2} m_{b,t}/v_{1,2}$). 
Then quartic scalar interaction parameters $\bar \lambda_i$, $i=$5,6,7
are affected by radiative corrections from one-loop diagrams with scalar
quarks of the order of $\mu^2 A^2/M^4_{SUSY}$, $\mu^3 A/M^4_{SUSY}$
and $\mu A^3/M^4_{SUSY}$ \cite{PilaftsisWagner}, so the phases of
$\bar \lambda_i$, $i=$5,6,7 are defined by the phases of complex $\mu$
and $A$, thus constraining the phase of $\mu^2_{12}$ in power of
the conditions (41),(43). 
In the case of the Born level MSSM potential with a global $U(1)_Q$
symmetry when $\bar \lambda_i=$0 
($i=$5,6,7) the complex $\mu^2_{12}$ parameter can still appear beyond the
tree-level due to the same CP violating Yukawa interactions of scalar 
quarks with the Higgs fields. This possibility of the $\mu^2_{12}$ phase 
induced in higher orders by radiative corrections calculated 
diagrammatically has been considered in
\cite{Pilaftsis}. 
The restoration of potential minimum can be 
achieved by means of the opposite sign quantum correction term, originating
from the tadpole diagrams with the pseudoscalar $A$ connected to the
squark loops 
\footnote{however, with nonzero $\lambda_5, \lambda_6$ and $\lambda_7$, 
the factor of the scalar-pseudoscalar Higgs counterterm is not explicitly
proportional to the tadpole renormalization constant, or the tadpole
parameter $c_0$.}.
  
In the classical minimum $c_0=$0 we find
\begin{eqnarray}
c_1&=&\frac{v^2}{2}( s_{\alpha} s^3_{\beta}-c_{\alpha} c^3_{\beta})
{\tt Im} {\bar \lambda_5}
+v^2 \, (s_{\alpha} c_{\beta} {\tt Im} {\bar \lambda_6} -
              c_{\alpha} s_{\beta} {\tt Im} {\bar \lambda_7})
\\ \nonumber
c_2&=&-\frac{v^2}{2}( s_{\alpha} c^3_{\beta}+c_{\alpha} s^3_{\beta})
{\tt Im} {\bar \lambda_5}
-v^2 \, (c_{\alpha} c_{\beta} {\tt Im} {\bar \lambda_6} +
              s_{\alpha} s_{\beta} {\tt Im} {\bar \lambda_7})
\nonumber
\end{eqnarray}
The second order terms $hA$
and $HA$ in (42) can be removed as usual by the orthogonal
rotation $a_{ij}$ ($i,j=$1,2,3) in $h,H,A$ sector 
\begin{eqnarray}
(h,H,A) \; M^2 \;  \left( \begin{array}{c} h\\ H\\ A 
\end{array} \right)
&=& (h_1, h_2, h_3) \; a^T_{ik} \; M^2_{kl} \; a_{lj} \; 
\left( \begin{array}{c} h_1 \\ h_2\\ h_3 
\end{array} \right) 
\end{eqnarray}
where the mass matrix has the form
\begin{eqnarray}
M^2 & = & \frac{1}{2} \left( \begin{array}{ccc}
m^2_h         &      0        & c_1 \\
0             &     m^2_H     & c_2 \\
 c_1          &     c_2       &     m^2_A
\end{array} \right) 
\end{eqnarray} 
Squared masses of the physical states $h_1, h_2, h_3$, which are the Higgs
bosons without definite CP-parity, are defined by the eigenvalues of
mass matrix $M^2$ 
(roots of the cubic equation for eigenvalues are given
by Cardano formulae)
\begin{eqnarray}
m^2_{h2} &=& 2 \sqrt(-q) {\tt cos} (\frac{\theta}{3})-\frac{a_2}{3} \\
\nonumber
m^2_{h1} &=& 2 \sqrt(-q) {\tt cos} (\frac{\theta+2\pi}{3})-\frac{a_2}{3}
\\ \nonumber
m^2_{h3} &=& 2 \sqrt(-q) {\tt cos} (\frac{\theta+4\pi}{3})-\frac{a_2}{3}  
 \nonumber
\end{eqnarray}
where
\begin{eqnarray*}
\theta&=&{\tt arccos} \frac{r}{\sqrt(-q^3)} \\
r=\frac{1}{54}(9 a_1 a_2 -27 a_0 - 2 a^3_2),&& q=\frac{1}{9}(3 a_1-a^2_2)
\\
a_0=c^2_1 m^2_H+c^2_2 m^2_h-m^2_h m^2_H m^2_A, &&
a_1=m^2_h m^2_H+m^2_h m^2_A+m^2_H m^2_A-c^2_1-c^2_2,\\
a_2=-m^2_h-m^2_H-m^2_A &&
\end{eqnarray*}
One can see that in the limiting case of CP-conserving potential
$c_{1,2}\to 0$ the following correspondence takes place: 
$m_{h_1} \to m_h$,
$m_{h_2} \to m_H$ and $m_{h_3} \to m_A$.
The normalized eigenvectors of the matrix $M^2$, which are at the
same time the matrix elements of $a_{ij}$, $(h,H,A)=a_{ij} h_j$,
have the form $a_{ij}=a^{'}_{ij}/n_j$, where 
\begin{eqnarray*}
a^{'}_{11}=((m^2_H-m^2_{h_1})(m^2_A-m^2_{h_1})-c^2_2), \;
a^{'}_{21}=c_1 c_2, \;
a^{'}_{31}=-c_1 (m^2_H-m^2_{h_1}) \\ 
a^{'}_{12}=c_1 c_2, \;
a^{'}_{22}=((m^2_h-m^2_{h_2})(m^2_A-m^2_{h_2})-c^2_1), \;
a^{'}_{32}=-c_2 (m^2_h-m^2_{h_2}), \\
a^{'}_{13}=-c_1 (m^2_H-m^2_{h_3}), \;
a^{'}_{23}=-c_2 (m^2_h-m^2_{h_3}),  \;
a^{'}_{33}=(m^2_h-m^2_{h_3})(m^2_H-m^2_{h_3})
\end{eqnarray*}
and $n_i=\sqrt(a^{'2}_{1i}+a^{'2}_{2i}+a^{'2}_{3i}$)
\footnote{No ordering of masses $m_{h_1}< m_{h_2}< m_{h_3}$
is required. If we want to keep this ordering, then
$a_{ij}$ written above, valid for the case $m_H> m_A$, must be changed. 
For the case
$m_H< m_A$ one should replace
$m_{h_2} \leftrightarrow m_{h_3}$  and change the sign of $a_{i2}, a_{i3}$
in the expressions for $a_{ij}$.}
.
Representations for the triple and quartic
Higgs boson self-interactions in the case of CP violating potential
are given by the expansions of structures $a_{ij} h_j a_{ik} h_k 
a_{il} h_l$, and $a_{ij} h_j a_{ik} h_k a_{il} h_l a_{i,m} h_m$, they are
bulky and not very telling, so we do not show them here. 
If the imaginary parts of $\bar \lambda_6$ and $\bar \lambda_7$
are not small, large off-diagonal elements of the mixing matrix $a_{ij}$ 
could appear leading to significant mass splittings of the Higgs states 
and modifications of the Higgs boson interactions.  

We assume that in the Yukawa sector $<\varphi_1>$
couples only to down fermions
\begin{equation}
\sum_{\alpha=d,s,b} V_{u \alpha} \frac{e m_\alpha}{2 \sqrt{2} m_W s_W
c_{\beta}}
       [\bar \psi_1 (1+\gamma_5) \psi_{2 \alpha} \varphi_1
        + \bar \psi_{2 \alpha} (1-\gamma_5) \psi_1 \varphi_1^+]
\end{equation}
where $(h,H,A)=a_{ij} h_j$, for the first generation quarks $\bar \psi_1= 
\{ \bar u, V_{ud} \bar d+V_{us}\bar s+ V_{ub} \bar b\}, \quad 
\psi_{2 \alpha}=(d,s,b)$ and analogous terms for $c$
and $t$ quarks ($V_{ab}$ denotes the CKM
matrix elements), and $<\varphi_2>$ couples only to up fermions
(model of type II \cite{typeII}):
\begin{equation}
    \frac{e m_u}{2 \sqrt{2} m_W s_W s_{\beta}}
       [\bar \psi_1 (1+\gamma_5) i \tau_2 \psi_2 \varphi_2^+
        + \bar \psi_2 (1-\gamma_5) i \tau_2 \psi_1 \varphi_2]
\end{equation}
where again physical $h_1, h_2, h_3$ states are introduced by means
of the $a_{ij}$ rotation, $\bar \psi_1= \{\bar u, V_{ud} \bar d +V_{us} 
\bar s+ V_{ub} \bar b\}, \quad \psi_2=u$ and analogous terms for $c$   
and $t$ quarks. 

\section{Higgs-gauge boson and Higgs-fermion couplings in the
MSSM with explicit CP violation}

In the following we shall focus on the MSSM scenario for the
two-Higgs-doublet model, which allows to restrict strongly the THDM
parameter space. It is not the only one possible, Standard Model-like 
scenarios in the general two-Higgs-doublet 
model have been discussed in \cite{osland}.
Detailed consideration in the framework of MSSM has been performed in
\cite{PilaftsisWagner} (also \cite{Ellis}). In this section 
we would like only to compare qualitatively our results with 
the results of these approaches.
Our calculation follows somewhat different scheme. In
\cite{PilaftsisWagner} the tree-level two-Higgs doublet potential
is CP invariant. The phase $\xi$ of $\mu^2_{12}$ is 
radiatively induced by the tadpole diagrams and can be 
absorbed in the definition of the $\mu$ parameter which 
appears in the stop mixing matrix off-diagonal element $A_t-\mu/{\tt 
tg}\beta$. The $\bar \lambda_5$, $\bar \lambda_6$ and $\bar \lambda_7$ 
terms are also radiatively induced by the threshold effects. At the 
same time the trilinear couplings $A_t$, $A_b$ also carry a phase
\footnote{ for a rewiev see e.g. \cite{cp_mssm}}
, so both the radiatively induced and the trilinear phases 
contribute to the phase {\tt arg}($\mu A$) of the $\bar \lambda_5$,
$\bar \lambda_6$ and $\bar \lambda_7$ terms.
We do not account for the radiatively induced phase
which is calculated diagrammatically.
In the case under consideration the fine-tuning conditions
(38),(39) are not fulfilled, so CP invariance of the 
two-doublet potential is explicitly broken by complex 
parameters at the tree level. The real and imaginary parts 
of $\mu^2_{12}$ parameter
are defined by means of the condition (41) for the real parts of 
$\bar \lambda_{5,6,7}$ parameters
and the minimization condition (43) (where $c_0=$0) for their
imaginary parts.
In the following calculations complex parameters $\bar \lambda_{5,6,7}$ 
are specified in the framework of the MSSM.

We used the two-loop symbolic results for 
$\bar \lambda_i$, $i=$1,...7, which were obtained 
in the RG approach \cite{twoloop} and
extended to the case of CP violation in \cite{PilaftsisWagner}.
The parameters $\bar \lambda_5$, $\bar \lambda_6$ and $\bar \lambda_7$
are nonzero in the next-to-leading order approximmation (RG improved
leading order approximation), so using (41) and (43)
\begin{eqnarray}
{\tt Re} \mu^2_{12}&=&
m^2_A s_{\beta} c_{\beta}+v^2 ( s_{\beta} c_{\beta}
{\tt Re} \bar \lambda_5 
+\frac{1}{2} c^2_{\beta} \; {\tt Re} \bar \lambda_6
+\frac{1}{2} s^2_{\beta} \; {\tt Re} \bar \lambda_7) \\
{\tt Im} \mu^2_{12}&=& 
\frac{v^2}{2} (  s_{\beta} c_{\beta} {\tt Im} \bar \lambda_5 +c^2_{\beta}
  \; {\tt Im} \bar \lambda_6 
+s^2_{\beta} \; {\tt Im} \bar \lambda_7)
\end{eqnarray}
where $\bar \lambda_{5,6,7}$ depend on the finite term 
corrections to the leading logarithmic result which appear from the 
one-loop diagrams with trilinear couplings. The analytical representation
of \cite{PilaftsisWagner} has the form
\begin{eqnarray}
\bar \lambda_5 &=& -\frac{3}{192\pi^2} h^4_t
                \frac{\mu^2 A^2_t}{M^4_{SUSY}}
                [1-\frac{1}{16\pi^2}(2 h^2_b -6 h^2_t
          +16g^2_s)t] \\ \nonumber
 &&     -\frac{3}{192\pi^2} h^4_b
                \frac{\mu^2 A^2_b}{M^4_{SUSY}}
                [1-\frac{1}{16\pi^2}(2 h^2_t -6 h^2_b
          +16g^2_s)t] \\ \nonumber
\bar \lambda_6 &=& \frac{3}{96\pi^2} h^4_t
                \frac{|\mu|^2 \mu A_t}{M^4_{SUSY}}
          [1-\frac{1}{16\pi^2}(\frac{7}{2} h^2_b -\frac{15}{2} h^2_t           
          +16g^2_s)t] \\  \nonumber
 &&  \hspace{-1cm}    -\frac{3}{96\pi^2} h^4_b                                 
                \frac{\mu}{M_{SUSY}}
             (\frac{6 A_b}{M_{SUSY}}-\frac{|A_b|^2 A_b}{M^3_{SUSY}})            
          [1-\frac{1}{16\pi^2}(\frac{1}{2} h^2_t -\frac{9}{2} h^2_b
          +16g^2_s)t] \\ \nonumber
\bar \lambda_7 &=& \frac{3}{96\pi^2} h^4_b
                \frac{|\mu|^2 \mu A_b}{M^4_{SUSY}}
          [1-\frac{1}{16\pi^2}(\frac{7}{2} h^2_t -\frac{15}{2} h^2_b
          +16g^2_s)t] \\ \nonumber
 && \hspace{-1cm}     -\frac{3}{96\pi^2} h^4_t   
                \frac{\mu}{M_{SUSY}}
             (\frac{6 A_t}{M_{SUSY}}-\frac{|A_t|^2 A_t}{M^3_{SUSY}})            
          [1-\frac{1}{16\pi^2}(\frac{1}{2} h^2_b -\frac{9}{2} h^2_t
          +16g^2_s)t] \nonumber
\end{eqnarray}
$A_{t,b}$ and $\mu$ are the factors in front of 
Higgs-squark(left)-squark(right) trilinear terms, $M_{SUSY}$ is the SUSY
energy scale,
$m_{top}$, $m_b$ are the on-shell running masses of the third generation 
quarks, 
and $t={\tt log}\frac{M^2_{SUSY}}{m^2_{top}}$,
$h_t=\frac{\sqrt{2} m_{top}}{v s_{\beta}}$,    
$h_b=\frac{\sqrt{2} m_{b}}{v c_{\beta}}$, $g_s=\sqrt{4 \pi \alpha_s}$.          
The trilinear parameters $A_t$, $A_b$ and $\mu$ of the Higgs boson
interaction with the third generation squarks can be generally speaking 
complex. In this case
the $\bar \lambda_i$, $i=$1,...7 parameters of the two-doublet Higgs
potential are defined by ${\tt tg}\beta$, SUSY scale $M_{SUSY}$,
and six relevant parameters in the sector of Higgs boson interaction
with the third generation squarks: $\mu$, arg($\mu$),
$A_t$, arg($A_t$), $A_b$, arg($A_b$). In the following consideration
for simplicity we assume $|A_t|=|A_b|$ and
assign the universal phase $\theta$ to $\mu A_t$ and $\mu A_b$ so that
$\theta=$arg($\mu A_t)=$arg$(\mu A_b)$. Then using the explicit structure
of (52) CP invariance conditions (39) can be rewritten in the form  
${\tt Im}(\stackrel{*} \mu^2_{12} \mu A)=$0 \cite{PilaftsisWagner}.

The couplings of $W$ and $Z$ bosons to the $h_1,h_2,h_3$ scalars have the 
form
\begin{center}
\begin{tabular}{lc}
$V_{\mu} \; V_{\nu} \; h_1$ & \hskip 1.5cm $f_V
                      g_{\mu \nu} (c_{\alpha-\beta}a_{21}
                                   -s_{\alpha-\beta} a_{11})$ \\
$V_{\mu} \; V_{\nu} \; h_2$ & \hskip 1.5cm $f_V
                      g_{\mu \nu} (c_{\alpha-\beta}a_{22}
                                   -s_{\alpha-\beta} a_{12})$ \\
$V_{\mu} \; V_{\nu} \; h_3$ & \hskip 1.5cm $f_V 
                      g_{\mu \nu} (c_{\alpha-\beta}a_{23}
                                   -s_{\alpha-\beta} a_{13})$ \\
\end{tabular}
\end{center}
where $V=W,Z$, $f_W=\frac{e}{s_W} m_W$ and
$f_Z=\frac{e}{s_W c^2_W} m_W$. The couplings of $h_1,h_2,h_3$
bosons to the $t$ and $b$ quarks have the form
\begin{center}
\begin{tabular}{lc}
$\bar t \; t \; h_1$ & \hskip 1.2cm $f_t \frac{1}{s_{\beta}}
    (s_{\alpha} a_{21}+c_{\alpha} a_{11}-i c_{\beta} a_{31} \gamma_5$) \\
$\bar t \; t \; h_2$ & \hskip 1.2cm $f_t \frac{1}{s_{\beta}}
    (s_{\alpha} a_{22}+c_{\alpha} a_{12}-i c_{\beta} a_{32} \gamma_5$) \\
$\bar t \; t \; h_3$ & \hskip 1.2cm $f_t \frac{1}{s_{\beta}}
    (s_{\alpha} a_{23}+c_{\alpha} a_{13}-i c_{\beta} a_{33} \gamma_5$) \\ 
& \\
$\bar b \; b \; h_1$ & \hskip 1.2cm $f_b \frac{1}{c_{\beta}}
    (c_{\alpha} a_{21}-s_{\alpha} a_{11}-i s_{\beta} a_{31} \gamma_5$) \\
$\bar b \; b \; h_2$ & \hskip 1.2cm $f_b \frac{1}{c_{\beta}}
    (c_{\alpha} a_{22}-s_{\alpha} a_{12}-i s_{\beta} a_{32} \gamma_5$) \\
$\bar b \; b \; h_3$ & \hskip 1.2cm $f_b \frac{1}{c_{\beta}}
    (c_{\alpha} a_{23}-s_{\alpha} a_{13}-i s_{\beta} a_{33} \gamma_5$) \\
\end{tabular}
\end{center}
where $f_{t,b}=-\frac{e}{2s_W} \frac{m_{t,b}}{m_W}$.

The Higgs boson mass
spectrum of the CP conserving limit $\theta=$0
(in this limit $a_{ij}=diag\{1,1,1\}$) is shown in Fig.1.
For the case of explicit CP violation in the two-doublet Higgs potential 
we take the parameter set $\mu=$-2 TeV, $A_t=A_b=$-1.8 TeV, 
$M_{SUSY}=$0.5 TeV, $m_A=$220 GeV, ${\tt tg}\beta=$4
which is typical for the region of MSSM parameter space where the 
imaginary parts of $\bar \lambda_5$, $\bar \lambda_6$ 
and $\bar \lambda_7$ are large (of the order of 0.1-1)
\footnote{A detailed discussion of possible combined constraints on the 
MSSM parameter space from the cosmology, direct searches and indirect 
measurements (rare decays) can be found in \cite{constraints}} .
We demonstrate in Fig.2 the neutral Higgs boson masses given by (47) and 
the mixing matrix elements $a_{ij}$ as a function of the 
universal phase $\theta=$arg($\mu A_{t,b})$. 
The Higgs boson masses of the CP conserving limit are
substantially changed when the phase $\theta$ is not small. The $m_{h1}$ 
in Fig.2 is always smaller than $m_h$ and $m_{h_2}$ has a downfall at 
the phase values around $\pi/4$.
The Higgs-vector boson $WWh_i$, $ZZh_i$ and 
the Higgs-fermion $q\bar q h_i$ ($q$=t,b) interaction vertices 
as a function of the phase $\theta$ are shown in Fig.3. One can observe 
that the $h_1$ couplings to gauge bosons $W,Z$ decrease by about 15\% 
if the phase of $\bar \lambda_5$, $\bar \lambda_6, \; \bar \lambda_7$ is 
large enough. Nonzero couplings of $h_3$ to gauge bosons appear.
The changes of the $b \bar b h_1$ and the $b \bar b h_2$ coupling
regime are also rather pronounced (see Fig.3).
In the region of MSSM parameter space where the $m_{h_3}$ is around
150-250 GeV and the $\mu$ and $A_{t,b}$ parameters are of the order 
of TeV the regime of strong mixing in the Higgs sector takes place. 
As a result the light Higgs boson $h_1$ could have not been
observed at LEP2 ($\sqrt{s}=$200 GeV) in the production channels
$e^+ e^- \to h_1 Z$, $e^+ e^-\to \nu_e \bar \nu_e h_1$ for the reason of 
suppressed couplings to gauge bosons, while the $h_2, h_3$ bosons 
are
sufficiently heavy to be not produced on mass-shell at the LEP2 energy. 
Detailed analysis of this scenario can be found in 
\cite{PilaftsisWagner,Ellis}. 

\section{Triple and quartic Higgs boson couplings in the MSSM
with explicit CP violation}

In the regions of the MSSM parameter space where the couplings
of lightest Higgs boson $h_1$ to gauge bosons and top quark are 
suppressed, traditional channels of Higgs boson production by radiation
from $W,Z$ or $t$ line and $WW,\, ZZ$ fusion can have too small rate
to be experimentally observed. For this reason it is interesting
to consider the possibility of double Higgs production 
(like $gg \to h_2\to h_1 h_1$) defined by the self-coupling vertices.
Such calculations are known in the CP conserving limit 
\cite{reconstruction}, when the cross sections of double and triple
Higgs boson production turn out to be very small. 
Only some of them are accessible for observation at high luminosity
colliders. In the case of CP violation some self-couplings can be
substantially increased, providing better possibilities for
the experimental reconstruction.
  
The $\bar \lambda_5$, $\bar \lambda_6$ and $\bar \lambda_7$ potential 
terms can modify significantly the Higgs boson self-interaction 
vertices calculated in the leading one-loop approximation with
$\bar \lambda_i=$0 ($i=$5,6,7).
At the  next-to-leading order approximation
the $\bar \lambda_i$ couplings (52) include the terms of the order of  
$h^4_{t,b} \, \mu^2 A^2_{t,b}/M^4_{SUSY}$ and $h^4_{t,b} \, \mu
A_{t,b}/M^2_{SUSY}$, so they can reach the values
of the order of 0.1-1 at moderate values of $M_{SUSY}$ and
$\mu$ and $A_{t,b}$ taken at TeV energy scale . For example, 
in the CP conserving limit $\theta=$0 the $hhh$ vertex 
in the mass parametrization has the form
\begin{eqnarray}
g_{hhh}&=&\frac{3e}{m_W s_W s_{2\beta}}
[-(c_{\beta} c^3_{\alpha}-s_{\beta} s^3_{\alpha})m^2_h
 +c^2_{\beta-\alpha} c_{\beta+\alpha} m^2_A  \\ \nonumber
&& +c^2_{\beta-\alpha}(\bar \lambda_5 c_{\beta+\alpha}
                    +\bar \lambda_6 c_{\beta} s_{\alpha}
                    -\bar \lambda_7 s_{\beta} c_{\alpha}) v^2]
\end{eqnarray}
The contributions of $\bar \lambda_{5,6,7}$ terms and the $m^2_{h,A}$ 
terms 
in this expression are of the same order if $\lambda_{5,6,7} \sim O(1)$. 
The rotation of the $\theta=$0 mass eigenstates by the matrix $a_{ij}$
defined by (45) gives the $g_{h_1 h_1 h_1}$ vertex of different form
but also with substantial contributions of the $\bar \lambda_{5,6,7}$ 
terms. 

Using the parameter set described in the previous
section, we show the values of various triple and quartic Higgs
boson self-interaction vertices as a function of the universal phase
$\theta=$arg($\mu A_t)=$arg($\mu A_b$) in Fig.4. The values of Higgs boson
self-interaction vertices in the CP-conserving limit $\theta=$0,$\pi$ and
in the leading order approximation $\bar \lambda_5=\bar \lambda_6= \bar 
\lambda_7=$0 are marked in Fig.4 by horisontal
arrows. The $\bar \lambda_5$, $\bar \lambda_6$ and $\bar \lambda_7$
potential terms induced in the next-to-leading order approximation
introduce very large corrections to the triple and quartic
self-interactions of Higgs bosons. In the region of the MSSM
parameter space under consideration the difference of the
leading order and the next-to-leading
order vertex factors can be several times in some ranges 
of the phase variation.

\section{Summary}

We demonstrated the tree-level equivalence of the two-Higgs-doublet 
model potentials (1) and (9), where CP-invariance can be 
explicitly broken by the $\lambda_6$ term in (1) or by the 
complex $\mu^2_{12}$, $\bar \lambda_5$ terms
in (9). The parameters $\lambda_i$ (i=1,...6) of (1) and
$\mu^2_1$, $\mu^2_2$, $\mu^2_{12}$, $\bar \lambda_i$ (i=1,...5) of (9)
are related by the equations (10). In the case of real 
parameters the diagonalization of 
potential (9) in the ground state can be performed by means 
of the substitutions (14)-(20) which express the $\bar \lambda_i$ and
$\mu^2_1$, $\mu^2_2$ parameters
through the Higgs boson masses $m_h$, $m_H$, $m_A$, $m_{H\pm}$,
the mixing angles $\alpha$, $\beta$ and the $\mu^2_{12}$ parameter. 
In the general case the $\bar \lambda_6$ and $\bar \lambda_7$ potential
terms (21) should be also considered with the diagonalization and 
minimization
conditions (22)-(28). If the complex parameters
$\mu^2_1$, $\mu^2_2$, $\bar \lambda_i$ (i=1,...7)
and $\mu^2_{12}$ 
are introduced, CP invariance of the hermitian potential (37)
is explicitly violated at the tree-level unless the fine-tuning conditions
(38) or (39) for parameters are satisfied. So in the following 
we consider the problem of diagonalization in the local minimum
for the two-Higgs-doublet potential which is not CP invariant.
For the diagonalization
of potential (37) again we use the substitution (22)-(28)
to be taken for real parts of parameters.
The minimization of potential (37)
at the tree level takes place with the condition $c_0=$0 (43)
for the imaginary parts of parameters.
The imaginary parts of $\bar \lambda_5$, $\bar \lambda_6$ and 
$\bar \lambda_7$ 
give rise to the CP-odd/CP-even Higgs boson off-diagonal terms,
which are removed by the orthogonal rotation in ($h,H,A$) space,
giving mass eigenstates $h_1,h_2,h_3$ without definite CP-parity 
and with the mass spectrum and couplings substantially different from the 
masses and couplings of CP-even and CP-odd states $h,H,A$, 
if the imaginary parts of parameters $\bar \lambda_5$,
$\bar \lambda_6$ and $\bar \lambda_7$ are sufficiently large. 

In the framework of MSSM the real parts of $\bar \lambda_i$
(i=1,...5) couplings are fixed at the SUSY energy scale by the conditions 
(29). Radiative corrections to the $\bar \lambda^{SUSY}_i$ (i=1,...7) 
couplings 
are generated at the $m_W$ energy scale. The 
equations (31)-(33) express the mixing angle $\alpha$ 
and masses of Higgs bosons in terms of the radiative
corrections to $\bar \lambda^{SUSY}_i$ (i=1,...7) couplings (e.g. given by 
the RG 
evolution). They are valid independently on the particular scheme
which is used for calculation of radiative corrections to the
$\bar \lambda^{SUSY}_i$ (i=1,...7).   

In the next-to-leading order approximation the complex
$\bar \lambda_5$, $\bar \lambda_6$ and $\bar \lambda_7$ 
parameters are generated by the soft CP violating Yukawa interactions 
of Higgs bosons with the scalar quarks. Using the results of 
\cite{PilaftsisWagner} 
we calculated the Higgs-gauge boson, Higgs-fermion and the Higgs
triple and quartic couplings for a representative MSSM parameter set,
when the off-diagonal elements of the Higgs boson mixing matrix
are large. The $\bar \lambda_5$, $\bar \lambda_6$ and $\bar \lambda_7$ 
parameters introduce
significant corrections to the Higgs self-interaction, even in the
case when their effects on the Higgs-gauge boson and Higgs-fermion
couplings are rather small. These corrections could rather strongly
(by one-two orders of magnitude in comparison with the case of 
CP conservation) enhance or suppress some channels of multiple Higgs 
boson production at next colliders, providing discriminative
tests of CP violation in the Higgs sector and improved feasibility
to reconstruct experimentally the Higgs potential. 

\begin{center}
{\large \bf Acknowledgements}
\end{center}

M.D. thanks very much F.~Boudjema, I.~Ginzburg, N.~Krasnikov and P.~Osland 
for useful discussions. The work was partially supported by
RFBR grant 01-02-16710, scientific program "Universities of Russia",
CERN-INTAS grant 99-0377 and INTAS grants 00-00313, 00-0679.

\newpage
\begin{center}
\begin{tabular}{|l|l|} 
\hline
Fields in the vertex & Variational derivative of Lagrangian by fields \\
\hline
$h \quad h \quad h$ & $\frac{3\, e}{M_W \, s_w \, s^2_{2\beta}} 
     [ -s_{2\beta}(c^3_{\alpha} c_{\beta}-s^3_{\alpha} s_{\beta}) m^2_h
       +2 c^2_{\alpha-\beta} c_{\alpha+\beta} \mu^2_{12}]$ \\
$H \quad H \quad H$ & $\frac{3\, e}{M_W s_w s^2_{2\beta}}
     [ -s_{2\beta}(c^3_{\alpha} s_{\beta}+s^3_{\alpha} c_{\beta}) m^2_H
       +2 s^2_{\alpha-\beta} s_{\alpha+\beta} \mu^2_{12}]$ \\
$H \quad H \quad h$ & $\frac{e \, s_{\alpha-\beta}}{2M_W \, s_w\,
s^2_{2\beta}}
     [ -(2m^2_H+m^2_h) s_{2\alpha} s_{2\beta} 
       + 4(3 s_{\alpha} c_{\alpha}+s_{\beta} c_{\beta}) \mu^2_{12}]$ \\
$H \quad h \quad h$ & $-\frac{e \, c_{\alpha-\beta}}{2M_W \, s_w \,
s^2_{2\beta}} 
     [ (m^2_H+2m^2_h) s_{2\alpha} s_{2\beta}
       - 4(3 s_{\alpha} c_{\alpha}-s_{\beta} c_{\beta}) \mu^2_{12}]$ \\
$H \quad A \quad A$ & $-\frac{e}{M_W \, s_w \, s^2_{2\beta}}
     [ s_{2\beta} (s_{\alpha} c^3_{\beta}+c_{\alpha} s^3_{\beta}) m^2_H
      + s^2_{2\beta} c_{\alpha-\beta} m^2_A -2 s_{\alpha+\beta}
\mu^2_{12}]$ \\
$h \quad A \quad A$ & $ \frac{e}{M_W \, s_w \, s^2_{2\beta}}
     [ s_{2\beta} (s_{\alpha} s^3_{\beta}-c_{\alpha} c^3_{\beta}) m^2_h
       +s^2_{2\beta} s_{\alpha-\beta} m^2_A +2 c_{\alpha+\beta} 
\mu^2_{12}]$ \\
$h \quad H^+ \quad H^-$ & $ \frac{e}{M_W \, s_w \, s^2_{2\beta}}
     [ s_{2\beta} (s_{\alpha} s^3_{\beta}-c_{\alpha} c^3_{\beta}) m^2_h
       +s^2_{2\beta} s_{\alpha-\beta} m^2_{H^\pm}
       + 2 c_{\alpha+\beta} \mu^2_{12}]$ \\
$H \quad H^+ \quad H^-$ & $ -\frac{e}{M_W \, s_w \, s^2_{2\beta}}
     [ s_{2\beta} (s_{\alpha} c^3_{\beta}+c_{\alpha} s^3_{\beta})
m^2_{H} + s^2_{2\beta} c_{\alpha-\beta} m^2_{H^\pm}
        - 2 s_{\alpha+\beta} \mu^2_{12}]$ \\
\hline
\end{tabular}
\end{center}
\begin{center}
 Table 1. Triple Higgs boson interaction vertices in the
general two-Higgs-doublet model, $\mu_{12}$ parametrisation.
\end{center}

\unitlength=1cm

\begin{center}
\begin{tabular}{|l|l|}
\hline
Fields in the vertex & Variational derivative of Lagrangian by fields \\
\hline
$h \quad h \quad h$ & $\frac{3\, e}{M_W \, s_w \, s_{2\beta}}
     [ -(c^3_{\alpha} c_{\beta}-s^3_{\alpha} s_{\beta}) m^2_h
       + c^2_{\alpha-\beta} c_{\alpha+\beta} (m^2_A+v^2 \lambda_5)]$ \\
$H \quad H \quad H$ & $\frac{3\, e}{M_W s_w s_{2\beta}}
     [ -(c^3_{\alpha} s_{\beta}+s^3_{\alpha} c_{\beta}) m^2_H
       + s^2_{\alpha-\beta} s_{\alpha+\beta} (m^2_A+v^2 \lambda_5)]$ \\
$H \quad H \quad h$ & $\frac{e \, s_{\alpha-\beta}}{2M_W \, s_w\,
s_{2\beta}}  
     [ -(2m^2_H+m^2_h) s_{2\alpha} 
       + 2(3 s_{\alpha} c_{\alpha}+s_{\beta} c_{\beta}) (m^2_A+v^2
\lambda_5)]$ \\
$H \quad h \quad h$ & $-\frac{e \, c_{\alpha-\beta}}{2M_W \, s_w \,
s_{2\beta}}      
     [ (m^2_H+2m^2_h) s_{2\alpha} 
       - 2(3 s_{\alpha} c_{\alpha}-s_{\beta} c_{\beta}) (m^2_A+v^2
\lambda_5)]$ \\
$H \quad A \quad A$ & $-\frac{e}{M_W \, s_w \, s_{2\beta}}
     [ (s_{\alpha} c^3_{\beta}+c_{\alpha} s^3_{\beta}) m^2_H
       + c_{2\beta} s_{\alpha-\beta} m^2_A -s_{\alpha+\beta} v^2
\lambda_5]$ \\
$h \quad A \quad A$ & $ \frac{e}{M_W \, s_w \, s_{2\beta}}
     [ (s_{\alpha} s^3_{\beta}-c_{\alpha} c^3_{\beta}) m^2_h
       + c_{2\beta} c_{\alpha-\beta} m^2_A + c_{\alpha+\beta} v^2
\lambda_5]$ \\
$h \quad H^+ \quad H^-$ & $ \frac{e}{M_W \, s_w \, s_{2\beta}}
     [ (s_{\alpha} s^3_{\beta}-c_{\alpha} c^3_{\beta}) m^2_h
       +s_{\alpha-\beta} m^2_{H^\pm}
       +  c_{\alpha+\beta} (m^2_A+v^2 \lambda_5)]$ \\
$H \quad H^+ \quad H^-$ & $ -\frac{e}{M_W \, s_w \, s_{2\beta}}
     [ (s_{\alpha} c^3_{\beta}+c_{\alpha} s^3_{\beta})
m^2_{H} +  c_{\alpha-\beta} m^2_{H^\pm}
        -  s_{\alpha+\beta} (m^2_A+v^2 \lambda_5)]$ \\
\hline
\end{tabular}
\end{center}   
\begin{center}
 Table 2. Triple Higgs boson interaction vertices in the
general two-Higgs-doublet model, $\lambda_5$ parametrisation.
\end{center}

\newpage

\begin{center}
\begin{table}
\begin{tabular}{|l|l|}
\hline
Fields in the vertex & Variational derivative of Lagrangian by fields \\
\hline
$h \quad h \quad h \quad h \quad$ &
$-\frac{3}{4} \frac{e^2}{M^2_W \, s^2_w \, s^3_{2\beta}}
[4 s_{2\beta}(c^3_{\alpha} c_{\beta}-s^3_{\alpha} s_{\beta})^2 m^2_h
+ s_{2\beta} s^2_{2\alpha} c^2_{\alpha-\beta} m^2_H -8 c^2_{\alpha-\beta}
c^2_{\alpha+\beta} \mu^2_{12}]$ \\
$H \quad H \quad H \quad H \quad$ &
$\frac{3}{4} \frac{e^2}{M^2_W \, s^2_w \, s^3_{2\beta}}
[-4 s_{2\beta}(c^3_{\alpha} s_{\beta}+s^3_{\alpha} c_{\beta})^2 m^2_H
+ s_{2\beta} s^2_{2\alpha} s^2_{\alpha-\beta} m^2_h +8 s^2_{\alpha-\beta}
s^2_{\alpha+\beta} \mu^2_{12}]$ \\
$A^0 \quad A^0 \quad A^0 \quad A^0$ & 
$-3 \frac{e^2}{M^2_W \, s^2_w \, s^3_{2\beta}}
[ s_{2\beta}(s_{\alpha} c^3_{\beta}+c_{\alpha} s^3_{\beta})^2 m^2_H
  s_{2\beta}(c_{\alpha} c^3_{\beta}-s_{\alpha} s^3_{\beta})^2 m^2_h
 -2 c^2_{2\beta} \mu^2_{12}]$ \\
$H \quad H \quad H \quad h$ & $-\frac{3}{4} \frac{e^2 s_{2\alpha}
s_{\alpha-\beta}}{M^2_W \, s^2_w \, s^3_{2\beta}}
[2 s_{2\beta} (c^3_{\alpha} s_{\beta}+s^3_{\alpha} c_{\beta}) m^2_H
 +s_{2\beta} s_{2\alpha} c_{\alpha-\beta} m^2_h 
 -4 s_{\alpha+\beta}] \mu^2_{12}$ \\
$H \quad h \quad h \quad h$ & $-\frac{3}{4} \frac{e^2 s_{2\alpha}
c_{\alpha-\beta}}{M^2_W \, s^2_w \, s^3_{2\beta}}
[2 s_{2\beta} (c^3_{\alpha} c_{\beta}-s^3_{\alpha} s_{\beta}) m^2_h
 +s_{2\beta} s_{2\alpha} s_{\alpha-\beta} m^2_H 
 -4 c_{\alpha+\beta}] \mu^2_{12}$ \\
$H \quad H \quad h \quad h$ & $-\frac{1}{4} \frac{e^2}{M^2_W \, s^2_w \,
s^3_{2\beta}} [-s_{2\beta} s_{2\alpha} 
(3 s_{2\alpha} s^2_{\alpha-\beta}-4 s_{\alpha-\beta}
c_{\alpha+\beta} -2 s_{\alpha+\beta} s_{\alpha-\beta} c_{\alpha-\beta} )
m^2_h $ \\
  & $
+s_{2\beta} s_{2\alpha} (s_{2\beta}+3 s_{2\alpha} s^2_{\alpha-\beta})
m^2_H -8(3 s^2_{\alpha} c^2_{\alpha}
- s^2_{\beta} c^2_{\beta}) \mu^2_{12}]$ \\
$H \quad H \quad A^0 \quad A^0$ & $ \frac{1}{4} \frac{e^2}{M^2_W \, s^2_w
\, s^3_{2\beta}} [-2 s_{2\beta} s_{2\alpha} s_{\alpha-\beta}
(c_{\alpha} c^3_{\beta}-s_{\alpha} s^3_{\beta}) m^2_h
 -2 s^3_{2\beta} c^2_{\alpha-\beta} m^2_A$
\\ 
   & $
- s_{2\beta} ( s_{2\alpha} s_{2\beta} +3 s^2_{\alpha-\beta} 
s^2_{\alpha+\beta}- s_{2\beta} s^2_{\alpha-\beta})
m^2_H
+ 4 (c^2_{2\beta} s^2_{\alpha-\beta}+s^2_{\alpha+\beta}) \mu^2_{12}]$ \\
$h \quad h \quad A^0 \quad A^0$ & $ \frac{1}{4} \frac{e^2}{M^2_W \, s^2_w
\, s^3_{2\beta}} [-s^2_{2\beta}(4 c_{2\beta} c_{2\alpha} +
3 s^2_{\alpha-\beta} s^2_{\alpha+\beta}+s^4_{\alpha-\beta}) m^2_h
-2 s^3_{2\beta} s^2_{\alpha-\beta} m^2_A $ \\
   & $
-2 s_{2\beta} s_{2\alpha} c_{\alpha-\beta} 
(s_{\alpha} c^3_{\beta}+c_{\alpha} s^3_{\beta}) m^2_H
+ 2(s^2_{2\beta} s^2_{\alpha-\beta}+4(c_{\alpha} c^3_{\beta}-s_{\alpha}
s^3_{\beta})^2) \mu^2_{12}]$ \\
$H \quad A^0 \quad A^0 \quad h$ & $\frac{1}{4} \frac{e^2}{M^2_W \, s^2_w
\, s^3_{2\beta}} [ -2 s_{2\beta} s_{2\alpha} c_{\alpha-\beta}
(c_{\alpha} c^3_{\beta}-s_{\alpha} s^3_{\beta}) m^2_h
+s^3_{2\beta} s_{\alpha-\beta} c_{\alpha-\beta} m^2_A$ \\
   & $
                   -2 s_{2\beta} s_{2\alpha} s_{\alpha-\beta}
(s_{\alpha} c^3_{\beta}+c_{\alpha} s^3_{\beta}) m^2_H
+ 2(2 s_{2\alpha} c_{2\beta} - s_{2\beta} s_{\alpha-\beta}
c_{\alpha-\beta}) \mu^2_{12}]$ \\  
$H^+ \quad H^+ \quad H^- \quad H^-$ & $-2 \frac{e^2}{M^2_W \, s^2_w
\, s^3_{2\beta}} [ s_{2\beta} (c_{\alpha} c^3_{\beta}-s_{\alpha}
s^3_{\beta}) m^2_h
 +s_{2\beta} (s_{\alpha} c^3_{\beta}+c_{\alpha} s^3_{\beta}) m^2_H
 -2 c^2_{2\beta} \mu^2_{12}]$ \\
$H^+ \quad H^- \quad A^0 \quad A^0$ & $- \frac{e^2}{M^2_W \, s^2_w
\, s^3_{2\beta}} [ s_{2\beta} (c_{\alpha} c^3_{\beta}-s_{\alpha}
s^3_{\beta}) m^2_h
 +s_{2\beta} (s_{\alpha} c^3_{\beta}+c_{\alpha} s^3_{\beta}) m^2_H
 -2 c^2_{2\beta} \mu^2_{12}]$ \\
$H^+ \quad H^- \quad h \quad h$ & $\frac{1}{4}\frac{e^2}{M^2_W \, s^2_w
\, s^3_{2\beta}} [ -s_{2\beta} (4 c_{2\alpha} c_{2\beta}
+ 3 s^2_{\alpha-\beta} s^2_{\alpha+\beta}+s^4_{\alpha-\beta} ) m^2_h
- 2 s^3_{2\beta} s^2_{\alpha-\beta} m^2_{H^{\pm}} $ \\
   & $
-2 s_{2\beta} s_{2\alpha} c_{\alpha-\beta}
(s_{\alpha} c^3_{\beta}+c_{\alpha} s^3_{\beta}) m^2_H
+2(s^2_{2\beta} s^2_{\alpha-\beta} +4(c_{\alpha} c^3_{\beta}-s_{\alpha}
s^3_{\beta})^2 \mu^2_{12}]$ \\
$H^+ \quad H^- \quad H \quad H$ & $\frac{1}{4}\frac{e^2}{M^2_W \, s^2_w
\, s^3_{2\beta}} [ -2 s_{2\beta} s_{2\alpha} s_{\alpha-\beta}
(c_{\alpha} c^3_{\beta}-s_{\alpha} s^3_{\beta}) m^2_h
- 2 s^3_{2\beta} c^2_{\alpha-\beta} m^2_{H^{\pm}} $ \\
   & $
+ s_{2\beta}( s_{2\alpha} s_{2\beta} 
-3 s^2_{\alpha-\beta} s^2_{\alpha+\beta}+s^4_{\alpha-\beta} ) m^2_H
+4 ( c^2_{2\beta} s^2_{\alpha-\beta}+s^2_{\alpha+\beta}) \mu^2_{12}]$ \\
$H \quad H^+ \quad H^- \quad h$ & $\frac{1}{2}\frac{e^2}{M^2_W \, s^2_w
\, s^3_{2\beta}} [-s_{2\beta} s_{2\alpha} c_{\alpha-\beta}
(c_{\alpha} c^3_{\beta}-s_{\alpha} s^3_{\beta}) m^2_h
+ s^3_{2\beta} s_{\alpha-\beta} c_{\alpha-\beta} m^2_{H^{\pm}} $ \\
   & $
- s_{2\beta} s_{2\alpha} s_{\alpha-\beta}
(c_{\alpha} s^3_{\beta}+s_{\alpha} c^3_{\beta}) m^2_H
+2(2 s_{2\alpha} c_{2\beta}-s^2_{2\beta} s_{\alpha-\beta}
c_{\alpha-\beta}) \mu^2_{12}]$ \\ \hline
\end{tabular}
\end{table}
\end{center}
\begin{center}
 Table 3. Quartic Higgs boson interaction vertices in the
general two-Higgs-doublet model, $\mu_{12}$ parametrisation.
\end{center}

\newpage

\begin{figure}[t]
\begin{center}
\begin{picture}(8.5,12)   
\put(-4,-1){\epsfxsize=17cm
         \epsfysize=16cm \leavevmode 
\epsfbox{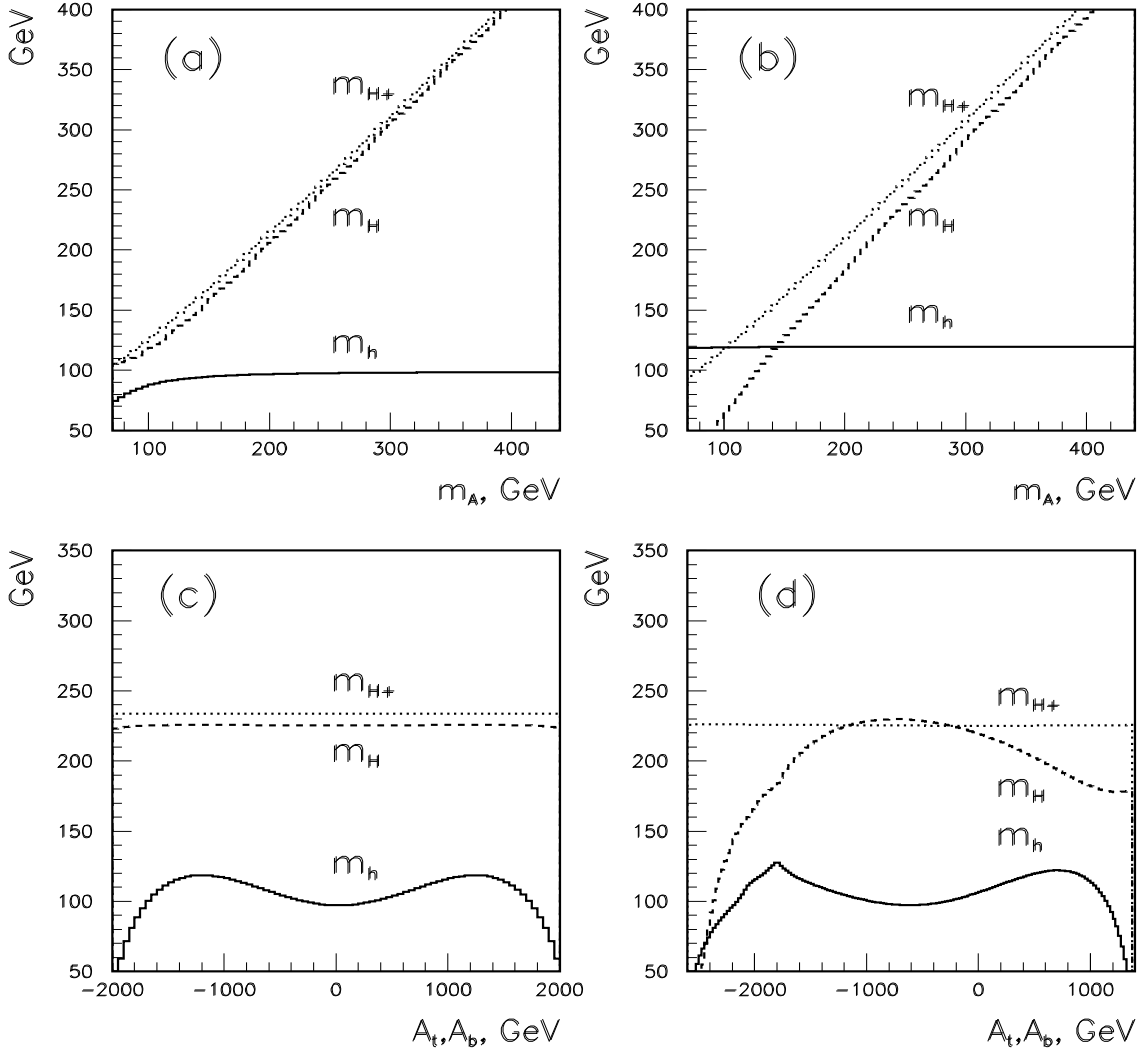}}
\end{picture}
\end{center}
\caption{Masses of the neutral and charged Higgs bosons 
$h,H,H^\pm$
versus the pseudoscalar mass $m_A$ and the trilinear constants 
$A_t$, $A_b$ calculated by means of (31)-(33) with
the analytical $\bar \lambda_i$ (i=1,...7) parametrization of 
\cite{PilaftsisWagner}. The $\Delta \bar \lambda_5$ is chosen to be
positive. The 
CP-conserving limit $\theta=$0 is taken. 
(a) ${\tt tg} \beta=$4, $M_{SUSY}=$0.5 TeV, $A_t=A_b=\mu=$0;
(b) ${\tt tg} \beta=$4, $M_{SUSY}=$0.5 TeV, $A_t=A_b=$0.9 TeV, $\mu=-1.5$ 
TeV; 
(c) ${\tt tg} \beta=$4, $M_{SUSY}=$0.5 TeV, $m_A=$220 GeV, $\mu=$0,
$A_t=A_b$;
(d) ${\tt tg} \beta=$4, $M_{SUSY}=$0.5 TeV, $m_A=$220 GeV, $\mu=-$2 TeV,
$A_t=A_b$.
Very small variations of the charged Higgs boson mass $m_{H^\pm}$
in (d) are due to the cancellation of leading power
terms $\sim \mu^2 A^2_{t,b}/M^4_{SUSY}$, see  \cite{PilaftsisWagner},
in the difference of $\Delta \bar \lambda_4$ and $\Delta \bar \lambda_5$, 
see (33).
If $\Delta \bar \lambda_5$ is chosen to be negative (purely imaginary $\mu 
A$
in (52)), $m_H$ increases in comparison with the case 
$A_t=A_b=\mu=$0.}
\end{figure}

\newpage

\begin{figure}[t]
\begin{center}
\begin{picture}(8.5,12)
\put(-3,0){\epsfxsize=15cm
         \epsfysize=14cm \leavevmode \epsfbox{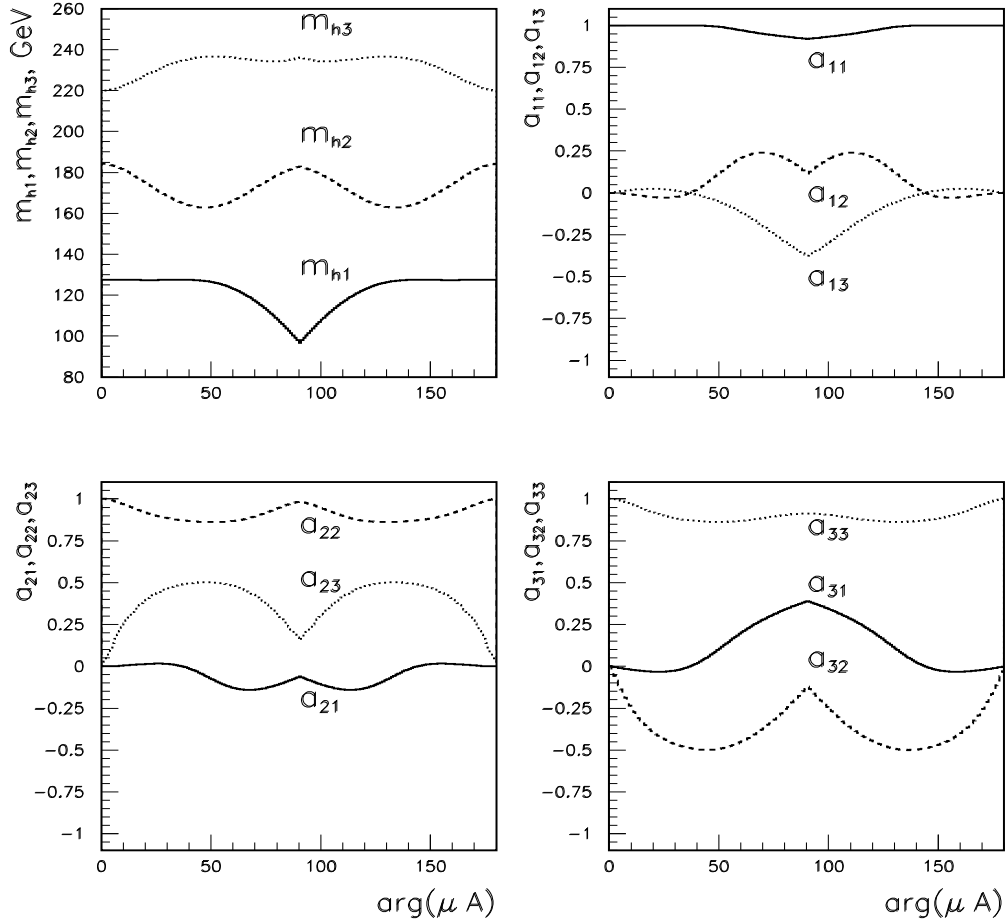}}
\end{picture}
\end{center}
\caption{Masses of the neutral Higgs bosons and the mixing matrix elements  
as a function of the $\bar \lambda_6$ and $\bar \lambda_7$ phase. The
$\bar \lambda_i$ parameters are taken from \cite{PilaftsisWagner} at the 
MSSM parameter values ${\tt tg}\beta=$4, $m_A=$220 GeV, $M_{SUSY}=$0.5 
TeV,
$A_t=A_b=-$1.8 TeV, $\mu=-$2 TeV.}
\end{figure}

\newpage

\begin{figure}[t]
\begin{center}
\begin{picture}(8.5,12)
\put(-3,0){\epsfxsize=15cm
         \epsfysize=14cm \leavevmode \epsfbox{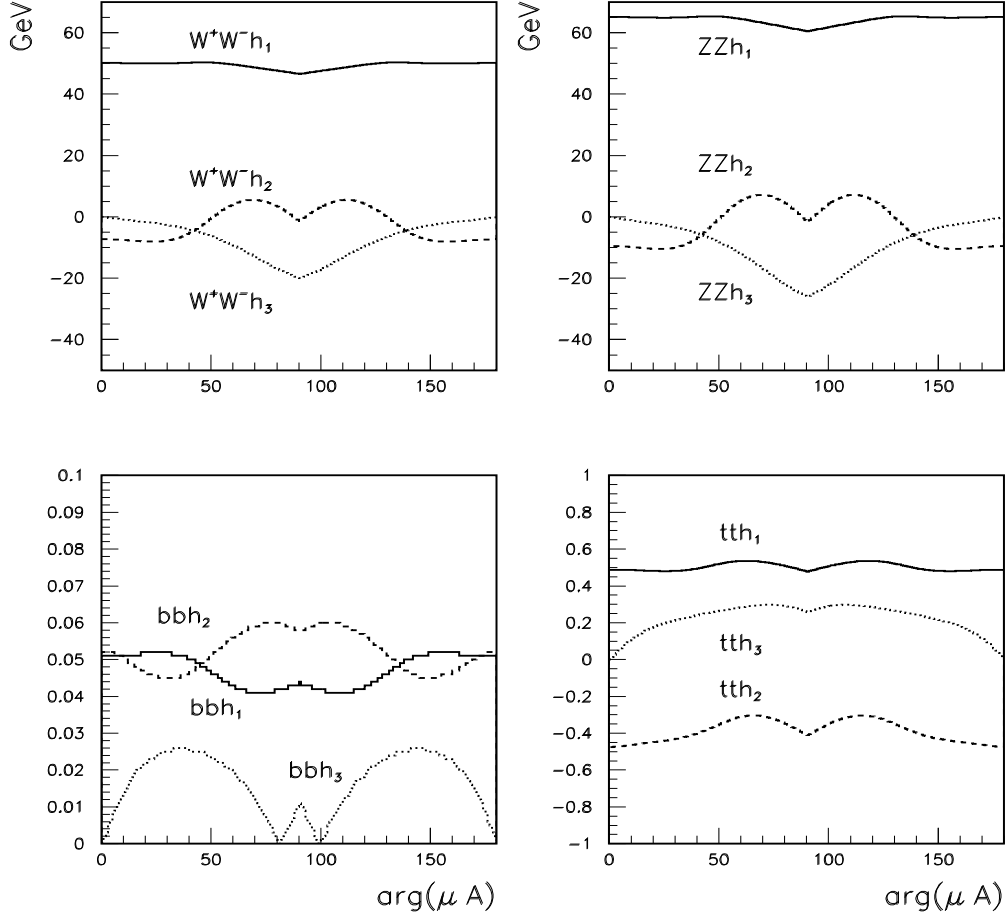}}
\end{picture}
\end{center}
\caption{Higgs-gauge boson and Higgs-fermion vertex factors  
as a function of the $\bar \lambda_6$ and $\bar \lambda_7$ phase. The 
$\bar \lambda_i$ parameters are taken from \cite{PilaftsisWagner} at the 
MSSM parameter values ${\tt tg}\beta=$4, $m_A=$220 GeV, $M_{SUSY}=$0.5 
TeV, 
$A_t=A_b=-$1.8 TeV, $\mu=-$2 TeV.  
For the coupling with fermions we plot 
$\sqrt{{\tt Im}^2 g_{ffh}+{\tt Re}^2 g_{ffh}}$.
}
\end{figure}

\newpage

\begin{figure}[t]
\begin{center}
\begin{picture}(8.5,12)
\put(-3,0){\epsfxsize=15cm
         \epsfysize=14cm \leavevmode \epsfbox{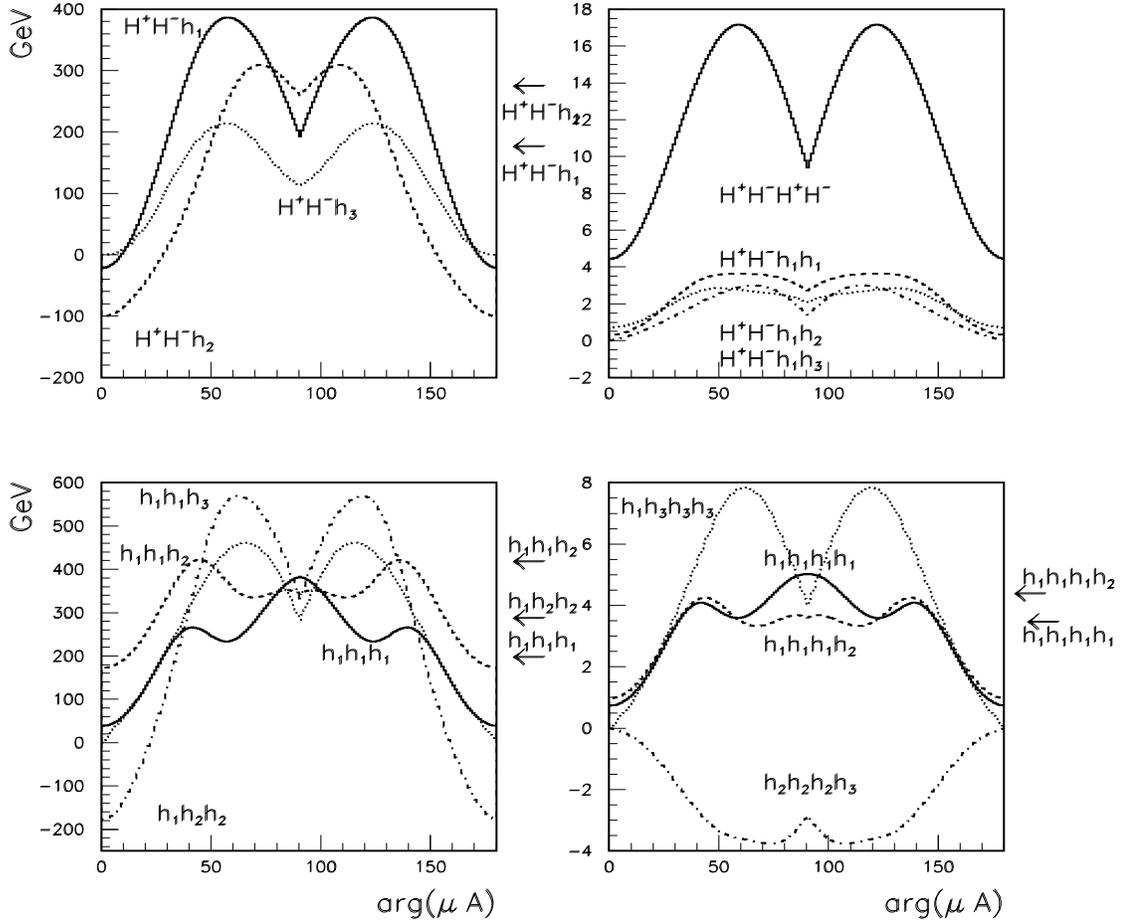}}
\end{picture}
\end{center}
\caption{Triple and quartic Higgs boson vertex factors
as a function of the $\bar \lambda_6$ and $\bar \lambda_7$ phase. The 
$\bar \lambda_i$ parameters are taken from \cite{PilaftsisWagner} at the 
MSSM parameter values ${\tt tg}\beta=$4, $m_A=$220 GeV, $M_{SUSY}=$0.5 
TeV, 
$A_t=A_b=-$1.8 TeV, $\mu=-$2 TeV.  
Horisontal arrows indicate the values of vertex factors in the
CP-conserving
limit $\theta=$0 and the leading order approximation $\bar \lambda_5= \bar 
\lambda_6= \bar 
\lambda_7=$0.}   
\end{figure}

\end{document}